\documentclass[epj,nopacs]{svjour}

\usepackage{graphicx,amssymb,amsmath,color}
\newcommand{\be}{\begin{equation}}
\newcommand{\ee}{\end{equation}}
\newcommand{\bea}{\begin{eqnarray}}
\newcommand{\eea}{\end{eqnarray}}
\newcommand{\br}{\mathbf{r}}

\def\eq#1{Equation~(\ref{#1})}

\usepackage[toc,page]{appendix}
\usepackage{array}
\newcolumntype{C}[1]{>{\centering\let\newline\\\arraybackslash\hspace{0pt}}m{#1}}
\newcolumntype{L}[1]{>{\raggedright\let\newline\\\arraybackslash\hspace{0pt}}m{#1}}
\newcolumntype{R}[1]{>{\raggedleft\let\newline\\\arraybackslash\hspace{0pt}}m{#1}}

\usepackage{subfigure}
\makeatletter
\renewcommand{\@thesubfigure}{\normalsize(\textbf{\alph{subfigure}})}

\begin{document}

\title{A Rationale for Mesoscopic Domain Formation~in~Biomembranes}

\author{Nicolas Destainville*, Manoel Manghi and  Julie Cornet}


\authorrunning{N. Destainville et al.}
\institute{Laboratoire de Physique Th\'eorique, IRSAMC, Universit\'e de Toulouse, CNRS, UPS, France, EU}

\abstract{Cell plasma membranes display a dramatically rich structural complexity characterized by functional sub-wavelength domains with specific lipid and protein composition. Under favorable experimental conditions, patterned morphologies can also be observed in vitro on model systems such as supported membranes or lipid vesicles. Lipid mixtures separating in liquid-ordered and liquid-disordered phases below a demixing temperature play a pivotal role in this context. Protein-protein and protein-lipid interactions also contribute to membrane shaping by promoting small domains or clusters. Such phase separations displaying characteristic length-scales falling in-between the nanoscopic, molecular scale on the one hand and the macroscopic scale on the other~hand, are named mesophases in soft condensed matter physics. In this review, we propose~a classification of the diverse mechanisms leading to mesophase separation in biomembranes. We~distinguish between mechanisms relying upon equilibrium thermodynamics and those involving out-of-equilibrium mechanisms, notably active membrane recycling. 
In equilibrium, we {especially focus on the many mechanisms that} dwell on an up-down symmetry breaking between the upper and lower bilayer leaflets. Symmetry breaking is an ubiquitous mechanism in condensed matter physics at the heart of several important phenomena. In the present case, it can be either spontaneous (domain buckling) or explicit, i.e., due to an external cause (global or local vesicle bending properties). Whenever possible, theoretical predictions and simulation results are confronted to experiments on model systems or living cells, which enables us to identify the most realistic mechanisms from a biological perspective.
}


\maketitle


%
%

%
%
%

\section{Introduction}
\label{intro}

The plasma membrane, a complex mixture of lipids and proteins, forms a selective barrier for eukaryotic cells~\cite{Alberts,PhillipsBook}, yet its role goes far beyond a simple frontier delimiting the cell interior~and~exterior. In the original 1972 model by Singer and Nicolson~\cite{Singer1972}, the plasma membrane was seen as a more or less homogeneous mixture in which the proteins represent about 50\% of the total mass. Since then, this model has known regular improvements, leading to the deciphering of an increasing organizational complexity, notably at the sub-micron level, and a growing understanding~of the biophysical and biochemical mechanisms at play~\cite{Edidin2003,Jacobson2007,Lenne2009,Komura2014,Jacobson2016,Destain2016,Sezgin2017}. There are evidences now that the membrane plays a crucial, active role in a large amount of biological functions~\cite{Alberts} such as viral and bacterial~infection, immune response, cell adhesion, transport of solutes or signaling,  \mbox{to name a few}, \mbox{and its organization} is directly related to its biological functions. The membrane is made up of~a lipid bilayer (mainly phospholipids, sphingolipids and cholesterol), the arrangement of which minimizes contact between water and the hydrophobic tails of these amphiphilic molecules~\cite{SafranBook}. The above-mentioned lipid mixture has been shown to present in eukaryotic cells two main distinct phases, termed liquid-disordered (Ld phase, cholesterol-poor) and liquid-ordered (Lo phase, cholesterol-rich)~\cite{Komura2014,Mouritsen2005,Marsh2009,Schmid2017} that undergo phase separation~\cite{Chaikin,Onuki2002,veatch1,Honerkamp2009} for a specific composition range and below $T\sim 20$--$30\ ^\circ$C depending on the nature of the lipid mixture, as it was already understood in the 1980's~\cite{Ipsen1987}.  The minimal requirement for such liquid phase coexistence seems to be a ternary mixture~of low- and high-melting temperature lipids, and cholesterol~\cite{Heberle2010}. 

The plasma membrane also contains various inclusions made of peripheral or integral proteins~\mbox{\cite{Alberts,Mouritsen2005}}. Thanks to recent in vivo and  in vitro experimental developments such as single particle tracking (SPT)~\cite{Kusumi2001,Daumas2003,Espenel2008}, fluorescence (or F\"orster) resonance
energy transfer (FRET)~\cite{Heberle2010,Feigenson2001}, atomic force microscopy (AFM)~\cite{Garcia2007,Grage2011,Connell2013,Whited2015,Ho2016}, super-resolution microscopy techniques (STED, PALM, STORM, SIM, see~the review~\cite{Lang2010}) or small-angle neutron scattering (SANS)~\cite{Heberle2013,Nickels2015,Usery2017}, it has been observed and it is now widely agreed that cell membrane components are heterogeneously distributed, and are generically organized into functional lipid and protein sub-micrometric or nanoscopic domains (nanodomains for short). In particular, a common view pictures lipid nanodomains as lipid ``rafts''~ 
\cite{Jacobson2007,Sezgin2017,Mouritsen2005,veatch1,Lingwood2010}, a consensual definition of which was formulated in 2006~\cite{pike}. They~are~dynamic supramolecular assemblies of nanometric length-scale (10 to 200~nm) enriched in sphingolipids and cholesterol (the above Lo phase) within a ``sea'' of Ld phase. 
 Specific proteins are then supposed to be targeted to them to perform their biological functions. However, the concept~of ``raft'' remains~controversial, especially in living cells~\cite{Schmid2017,veatch1,Lingwood2010,Heerklotz2002,Munro2003,Harder2003,Hancock2006,Poveda2008,Leslie2011,Klotzsch2013,Honigmann2014,Levental2016}. In particular the mechanisms for membrane rafts being so small are still debated. Indeed, creating many small domains instead of a single large one generates a longer boundary between Lo and Ld phases, resulting in an increased interfacial energy. Even their existence as just being the consequence of a lipidic meso-phase separation, independently~of proteins interacting with lipids and other proteins, is debated. Nevertheless, at~present, it can safely be concluded that most membrane proteins, as well as lipids, segregate into different domains via specific mechanisms. All these domains have proven to be key players in the above-mentioned biological functions, but a full and consensual understanding of their biophysical and biochemical origin~is still~lacking.

The aim of this review is precisely to expose the different models that have been proposed from~a physical perspective and to deliver a rationale to better understand the mesophase formation and stability on length-scales of biological interest. In the biophysical literature, the observed domains below the diffraction limit are often named ``nanodomains'' because their size ranges from tens~to hundreds of nanometers. In what follows, (nano)domains will refer either to protein domains (also~known as ``protein clusters''~\cite{Lang2010}), 
or to lipid domains with a composition different from the membrane bulk, or to lipid domains with a phase state different from the membrane bulk \mbox{(e.g., Lo domains in a Ld phase)}, or (and most likely) to any combination of these three situations. Indeed, most continuous theories that will be under consideration in this review are sufficiently general to embrace such a large spectrum of physical realities~\cite{Chaikin,Seul1995,Giuliani2008}. When needed, we~will specify which system is considered in particular. Apart from cell biology, alike membrane patterned morphologies can even be observed in different contexts of biology, such as pollen grains, \mbox{fungal~spores~or insect eggs}~\cite{Lavrentovich2016}. 

Two major classes of processes stand out and will be addressed here: mechanisms considering the membrane in thermodynamic equilibrium, and active or non-equilibrium processes, which integrate the dynamical interactions of the membrane with the cytosol, especially active membrane recycling. 

In thermodynamic equilibrium, different models, {many} of them being shown in this review~to~be based upon the symmetry-breaking principle, provide explanations for small domains formation.  {In this respect}, two main classes of mechanisms can be distinguished: spontaneous buckling and bending-related phenomena. Lipowsky indeed described in 1993 that membrane domains can arise due to a spontaneous symmetry breaking mechanism~\cite{lipowsky}. When the system~is quenched below the phase-separation temperature, domains coarsen, which increases their perimeter and then the interfacial energy. They eventually bend undergoing buckling in the third dimension \mbox{(perpendicular to the} membrane plane), so that the interface energy cost between the two phases~is~lowered. We will discuss into detail below that they are then trapped in a metastable state.

A second category of equilibrium models (and experiments) show that domain formation can rely on the coupling between curvature and local composition. The global frame of this type of mechanisms was underlined by Seul and Andelman in 1995~\cite{Seul1995}, bringing into play a competition between short-range attraction (leading to a finite line tension at interfaces between different phases) and longer-range repulsion, for example generated by electrostatic (e.g., dipole-dipole) interactions between lipids or proteins, and/or by their coupling to the membrane curvature as in the present~case. A typical patterning length-scale emerges set by the relative strengths and ranges of attraction and repulsion~\cite{Giuliani2008}, and can be tuned by varying control parameters such as temperature, membrane tension~or constituent concentrations. Generally speaking, local curvature occurs whenever the up-down symmetry is locally broken due to the presence of different lipid compositions in the two leaflets and/or non-symmetric proteins, or different types of solutes that can adsorb on the two sides~of the membrane. By ``broken up-down symmetry'', we mean that for a variety of possible reasons that will be discussed further in this Review, the upper and lower leaflets of the bilayer, or equivalently the inner and outer leaflets of a vesicle or cell, do not play the same role. 

A third type of membrane symmetry breaking in equilibrium is related to the presence of lipid species with different numbers of unsaturations, leading to thicker domains (often enriched with cholesterol, such as Lo phases) with a higher bending modulus inducing hydrophobic~mismatch. This~results in total demixing in a plane bilayer below the phase-separation temperature. But,~\mbox{on a curved} surface like a vesicle, the larger these domains are, the more difficult to bend they~are. The~minimization of the bending energy cost then favors the division of large thick domains into smaller ones that better accommodate the membrane to the global spherical shape~\cite{Goh2013}.

In this review, we will not consider in depth the potential role of so-called ``linactants'' (or~line-active molecules), the two-dimensional equivalent of surfactants~\cite{SafranBook}. Such molecules (proteins~or hybrid lipids having one saturated and one unsaturated hydrocarbon chain) localize~at the interface between coexisting phases and lower the interface line tension. They potentially stabilize patterned structures and micro-emulsions, which are actually strongly related to the concept of rafts~\cite{Mouritsen2005}. The interested reader can refer to the Reviews \cite{Komura2014,Schmid2017,Honerkamp2009,Palmieri2014} and to our Discussion Section. 

On the other hand, non-equilibrium models describe how active membrane recycling enters in competition with domain growth and then modulates their size in the stationary regime. Perpetual recycling of the cell plasma membrane due to material exchange with the cytosol, due for example to exocytosis and endocytosis of membrane patches, continually mixes the plasma membrane components and breaks too large membrane domains.

This review is organized as follows. After an introduction, Section~\ref{equil} puts into perspective the principal mechanisms in equilibrium accounting for mesophase separation in biphasic planar membranes or vesicles. The mechanisms deal with either separation of lipidic phases such as Lo and Ld ones, or condensation in domains of interacting proteins  embedded in lipid bilayer.  We~make the distinction between the so-called ``weak segregation limit'' in the vicinity of the miscibility critical point where the field theory at Gaussian order is often used. In principle, field~theories~at Gaussian order are only valid above the critical temperature and far from it~\cite{Chaikin}. However, applying them close~to the critical point amounts to use mean-field theories that provide reasonable orders of magnitude~of critical temperatures and critical exponents that are very useful as a first step. \mbox{To~go beyond} these simple estimates, far more involved renormalization techniques must be used, beyond the scope~of this~review.  
The ``strong segregation limit'', well below the critical temperature, where effective models using the notion of line tension are more efficient. We stress how up-down symmetry breaking~is then involved in the destabilization of the macrophase in favor of the mesophase. In~combination with experiments, the theoretical tools to tackle these questions go from all-atom and coarse-grained molecular dynamics~\cite{Chavent2016} and Monte Carlo simulations~\cite{Binder1986}, to analytical modeling, notably using a continuous, field-theoretic description of the membrane~\cite{Seul1995}. Section~\ref{active:memb} addresses the characterization of out-of-equilibrium membrane steady-states. Using tools from out-of-equilibrium statistical mechanics, either numerical or theoretical, the reviewed studies predict domain-size distributions that will have to be confronted to experiments in the future. This will enable biologists and biophysicists to identify the mechanisms that are actually at play in living cells. Finally, a discussion and conclusion Section discusses the different models from a cell biology perspective and proposes possible routes to be followed in future studies. Table~\ref{tab:notations} gives an overview of the principal notations used throughout the review. Notably, the natural energy scale will turn out to be the thermal energy $k_{\rm B}T\simeq 4.3 \times 10^{-21}$~J at physiological temperature. 

\begin{table}
\begin{center}
\begin{tabular}{|c|l|c|c|}
\hline
Notation & Name & Sec. defined & Refs. \\
\hline
Lo & Liquid-ordered lipid phase &\ref{intro} & \cite{Komura2014,Mouritsen2005,Marsh2009} \\
Ld & Liquid-disordered lipid phase &\ref{intro}  & \cite{Komura2014,Mouritsen2005,Marsh2009} \\
$\sigma$ & Membrane surface tension (i.e., energy per unit area) & \ref{equil}  & \cite{PhillipsBook,SafranBook} \\
$\kappa$ & Bending elastic modulus or curvature rigidity & \ref{equil} & \cite{PhillipsBook,SafranBook,Chaikin} \\
$\kappa_G$ & Saddle-splay elastic modulus &  \ref{equil} &  \cite{SafranBook,Chaikin} \\
$\xi$ & Helfrich correlation length: $\xi=\sqrt{\kappa/\sigma}$ & \ref{curv:comp:cp} & -- \\
$h(\mathbf{r})$ & Height function (in the Monge representation)&  \ref{equil} &  \cite{SafranBook} \\
$\phi(\mathbf{r})$ & Order parameter: local area fraction or phase state (e.g. Lo/Ld) &  \ref{equil} &  \cite{Mouritsen2005} \\
$H$ & Mean curvature ($H=\nabla^2 h/2$ in the Monge representation) &  \ref{equil} &  \cite{PhillipsBook,SafranBook,Chaikin} \\
$K$ & Gaussian curvature &  \ref{equil} &  \cite{PhillipsBook,SafranBook,Chaikin} \\
$v$ & Reduced volume of a vesicle of volume $V$ and area $A$: $v = 6 \sqrt{\pi} V/A^{3/2} \leq 1$ & \ref{buckling} & \cite{Hu2011}  \\
$C_{\rm sp}$ & Spontaneous or preferred mean curvature &  \ref{equil} &  \cite{SafranBook,Chaikin} \\
$\lambda$ & Domain line tension (i.e., energy per unit length at a 1D phase boundary) &  \ref{equil} &  \cite{Honerkamp2009,lipowsky} \\
$\xi_I$ & Invagination length: $\xi_I=\kappa/\lambda$ & \ref{buckling} & \cite{lipowsky} \\
$\Lambda$ or $\kappa C_1$ & Coupling coefficient between $\phi(\mathbf{r})$ and $H$ & \ref{equil} &  \cite{Honerkamp2009,Seul1995} \\
$T_c$ & Critical temperature (at a miscibility critical point) &  \ref{equil} &  \cite{SafranBook,Mouritsen2005,veatch1,Honerkamp2009} \\
$T_d$ & Demixing or phase-separation temperature &  \ref{equil} &  \cite{Mouritsen2005,veatch1,Honerkamp2009} \\
$\xi_{\rm OZ}$& Ornstein-Zernike composition correlation length &  \ref{equil}  & \cite{Chaikin,Barrat_book}.\\
$k_{\rm B} T$ & Thermal energy $\simeq 4.3 \times 10^{-21}$~J at physiological temperature (37$^{\rm o}$C) &\ref{intro} & \cite{PhillipsBook} \\
$\Delta p$ & Pressure jump across the membrane (for closed vesicles): $\Delta p=p_{\rm int}-p_{\rm out}$ &  \ref{vesi} &  \cite{helfrich1973}  \\
$\tau$ & Recycling time (out-of-equilibrium membranes) & \ref{models} & \cite{Foret2005,Turner2005} \\
\hline
\end{tabular}
\end{center}
\caption{Main notations used in this Review, together with the sections where they are defined and useful references. 
\label{tab:notations}}
\end{table}

\section{In Thermodynamic Equilibrium}
\label{equil}

Two-dimensional patterning has been known for decades in condensed matter physics and~its origin can generically be accounted for by a ``competing-interactions model'', as~described by Seul and Andelman in their seminal paper~\cite{Seul1995}. In their article, mesophases or ``modulated phases'' are \mbox{described in a} variety of contexts, such as semiconductor surfaces, superconductor~films, liquid~crystal~films, ferrofluids, polymer mixtures, diblock copolymers, Langmuir films, \mbox{and biphasic biomembranes}. 

The first theoretical works on the formation of curved meso-structures in membranes have been done by Leibler~\cite{leibler1} and Leibler and Andelman~\cite{leibler2} in the 1980's by adopting a field-theoretic~approach. They showed how coupling the membrane composition to its mean curvature can lead to modulated~phases.

At the mean-field level the system made of two types of molecules A and B is 
 described~as~a function of the thermodynamic variable $\phi$ which is the area fraction of molecules of type A, and~therefore $1-\phi$ is 
 the one of molecules of type $B$. 

If the coupling between composition and membrane height fluctuations is neglected, a~second-order macro-phase transition occurs between a disordered phase where lipids A and B are mixed and an ordered phase where two phases, the A-rich and B-rich ones (alternatively Lo and Ld phases), coexist. The critical area fraction is $\phi_c=1/2$ 
and the critical temperature, noted $T_c$, is~related~to the Flory interaction parameter $\chi\propto (\alpha_A-\alpha_B)^2$ where $\alpha_{A,B}$ are the electronic polarisabilities~of molecules A and B in this context. Close to $T_c$, the grand potential per unit area can be expanded in powers of $\psi=\phi-1/2$~\cite{Chaikin}
\bea
w(T,\mu,\psi)&=&f(\psi)-\mu\psi \\
&=&w(T,\mu,0)-h\psi+\frac{m}2 \psi^2-s \psi^3+u\psi^4+\mathcal{O}(\psi^5) 
\label{w}
\eea
where $f(\psi)$ is the free energy and $\mu$ the chemical potential (both per unit area). The coefficients in the expansion, $m=\alpha(T-T_c)$ (with $\alpha>0$) and $s$ are proportional to $T-T_c$, whereas $u$ is almost independent on $T$ and $h=\mu-\mu_c-\gamma(T-T_c)$. In particular the equation of state is given by $\partial w/\partial \psi=\partial f/\partial \psi -\mu=0$, and the coexistence curve in the $(\psi, T)$ plane can be drawn \mbox{close to $T_c$}. Note that if the area fraction is different form the critical value  1/2, the temperature at which the coexistence occurs is smaller and noted $T_d(\phi)<T_c$~\cite{Chaikin}. It is called the ``phase-separation'' or ``demixing'' temperature. 

To take into account the spatially varying composition $\phi(\br)$, one constructs the phenomenological Landau-Ginzburg Hamiltonian where a term in $\frac{b}2(\nabla \phi)^2$ is introduced, characterizing the energy cost associated with local variations of the concentration. The coefficient $b$ is related to the pair interaction between molecules A and B. Below $T_d$, the system phase-separates into two phases and an interface sets up. Minimizing this Hamiltonian (with $m$ negative, and $s=0$ which can always be fixed by shifting $\phi$) leads to the classical interfacial tension~\cite{SafranBook,Cahn_Hilliard}, in this 2D case a line tension $\lambda$~\cite{Honerkamp2009,lipowsky}, with~$\lambda \propto  b\Delta\phi^2/\xi_{\rm OZ} \propto |m|^{3/2}\sqrt{b}/u$ where $\Delta\phi=\sqrt{|m|/u}$ is the difference in composition of the two phases and $\xi_{\rm OZ}=\sqrt{b/|m|}$ is the Ornstein-Zernike correlation length of composition fluctuations in the bulk and far from the critical point~\cite{Chaikin,Barrat_book}.

\begin{figure}[h]
\begin{center}
 \subfigure[]{\includegraphics[width=7.5cm]{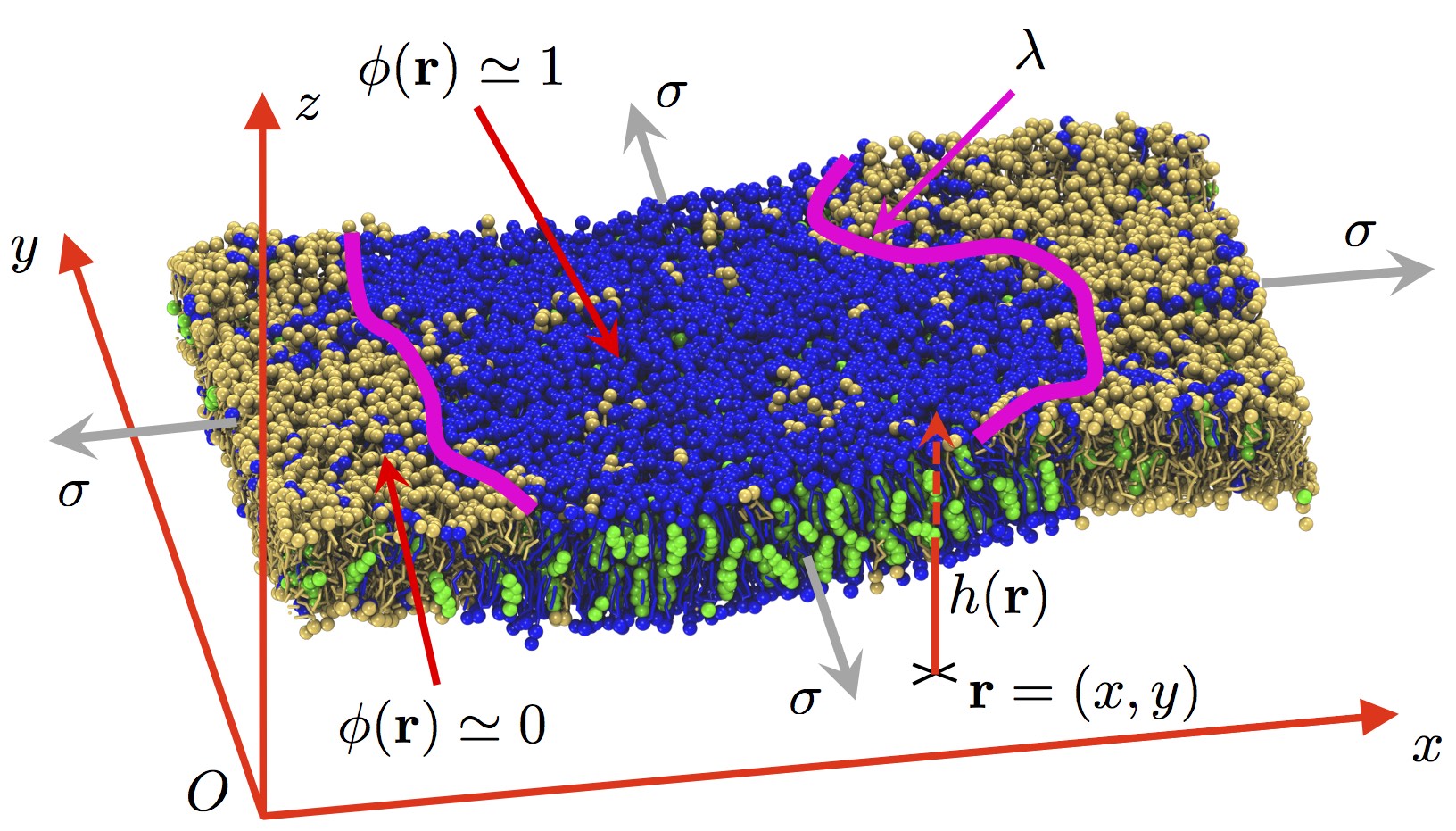}}
 \subfigure[]{\includegraphics[width=5.5cm]{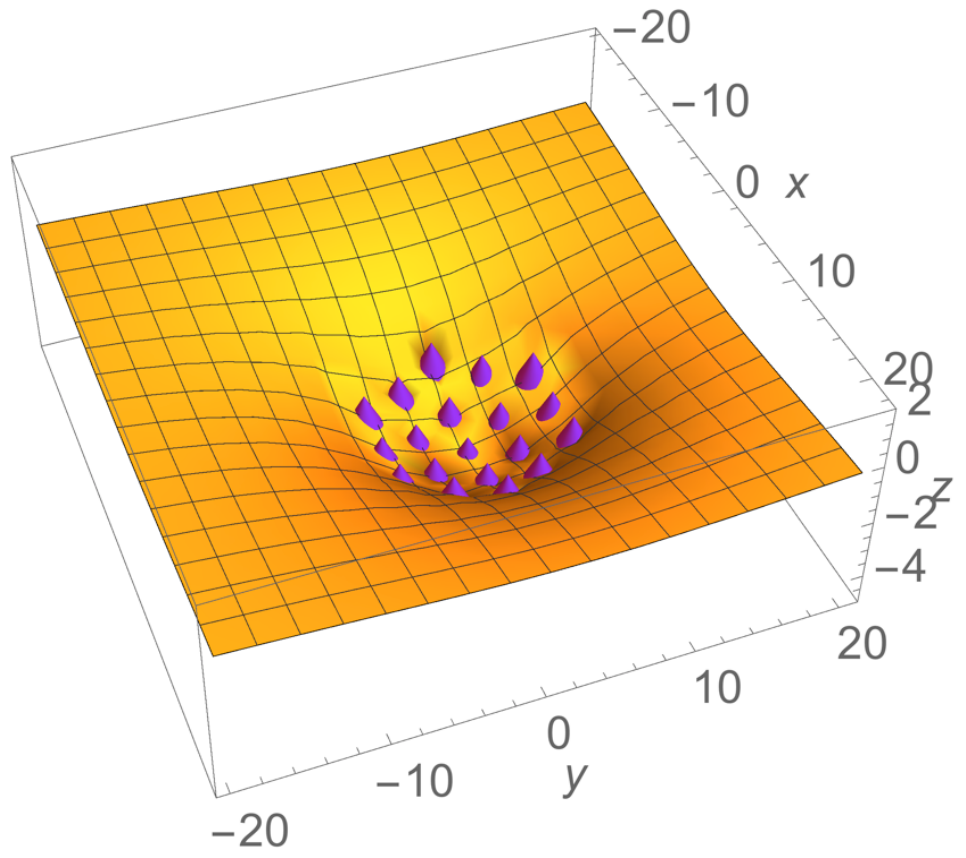}}
\end{center}
\caption{(\textbf{a})~Membrane simulation snapshot. The blue lipids are 1,2-dipalmitoylphosphatidylcholine (DPPC), the yellow ones di-C16:2-C18:2 PC (DIPC) 
 and cholesterol appears in green. The pink thick line is the 1D interface delimiting the liquid-ordered (Lo) phase (cholesterol rich, $\phi(\mathbf{r})\simeq 1$) and the liquid-disordered (Ld) phase 
 (cholesterol poor, $\phi(\mathbf{r})\simeq0$). This interface has an energy cost per unit length, homogeneous to a force $\lambda$, named the line tension. Although fluctuating, the membrane is globally planar and parallel to the plane $(xOy)$. The height of the membrane above this reference plane $(xOy)$ is measured by the height function $z=h(\mathbf{r})$. The membrane is taut with surface tension $\sigma$ as illustrated by the four gray arrows parallel to the plane $(xOy)$. Adapted from an original membrane image generated by the MARTINI 
 force field, reproduced with the courtesy of Matthieu Chavent. (\textbf{b})~In the cluster-phase scenario, proteins are described as individual objects embedded in a continuous fluctuating 2D lipid mattress, also represented by a height function. They gather because of  short-range attractive forces but long-range repulsion between clusters limits their growth. In the present case, each individual protein (in purple) locally imposes a spontaneous curvature to the membrane that is represented as an elastic sheet (in orange). {In a taut membrane, an effective long-range repulsion between proteins ensues (see text)}. The units are arbitrary.}
\label{two:fields}
\end{figure}

However, the lipids do not evolve on a stiff planar surface but on a fluctuating elastic membrane. In the case of an homogeneous lipid mixture, Helfrich proposed in 1973 that the elastic energy of a membrane or vesicle is given by the integral~\cite{PhillipsBook,SafranBook,helfrich1973} 
\begin{equation}
H_{\rm Helfrich} = \frac{\kappa}2 \int_\mathcal{A} {\rm d}A (c_1+c_2-2C_{\rm sp})^2 + \kappa_G \int_\mathcal{A} {\rm d}A c_1c_2 + \sigma \int_\mathcal{A} {\rm d}A
\label{Helf}
\end{equation}
over the whole membrane surface $\mathcal{A}$. 
Here $H=(c_1+c_2)/2$ is the local mean curvature, $K=c_1c_2$ the Gaussian curvature, and $C_{\rm sp}$ the spontaneous mean curvature. It measures the more or less pronounced membrane tendency to bend spontaneously, upward or downward. The bending modulus $\kappa$ typically falls  in the 10 to $100 \ k_{\rm B}T$ interval~\cite{Mouritsen2005,Schmid2017,Dimova2014}. In this respect, thicker Lo phases are known to have higher bending rigidities that Ld phases. For example, measured ratios are $\kappa_{\rm Lo}/\kappa_{\rm Ld} \simeq 5 $ in reference~\cite{Baumgart2005} and $\simeq 4$ in reference~\cite{Semrau2009}, a range of values confirmed in more recent studies measuring ratios up to 10~\cite{Nickels2015,Usery2017}. Membrane stability requires that $-2 < \kappa_G/\kappa < 0$ (see, e.g.,~\cite{Gutlederer2009}). The membrane surface tension~is denoted by $\sigma$ (see Figure~\ref{two:fields}a). Here $\sigma$ appears as a Lagrange multiplier controlling the membrane~area. There are several alternative definitions of the surface tension that coincide in the high tension limit~\cite{Gueguen2017}. It is imposed by external constraints, and cannot exceed the so-called ``lysis''~tension, on~the order of $10^{-2}$~N/m for usual lipids such as 1,2-Dioleoyl-sn-glycero-3-phosphocholine (DOPC).

In the case where the membrane adopts a globally planar configuration, it can be described by a height function $h(\br)$ measuring the distance to a reference plane, as in Figure~\ref{two:fields}a. Then $H \simeq \nabla^2 h/2$ when the fluctuations of $h$ remain small. The coupling between local composition and membrane height fluctuations was introduced by Leibler and Andelman in the following effective field-theoretic Hamiltonian (by skipping the irrelevant constant and ignoring the Gaussian curvature at this stage):
\begin{equation}
H[\phi,h]= \int_S {\rm d}^2 \mathbf{r} \left[w(\phi)+\frac{b}2(\nabla \phi)^2\right] + \frac12 \int_S {\rm d}^2 \mathbf{r} \left[    
\sigma (\nabla h)^2 + \kappa (\nabla^2 h)^2 \right] + \Lambda  \int_S {\rm d}^2 \mathbf{r} \, \phi(\mathbf{r}) \nabla^2 h (\mathbf{r})
\label{couplingLA}
\end{equation}
where $S$ is now the projected area in the $(xOy)$ plane. The constant $\Lambda$ couples the composition $\phi(\mathbf{r})$ and the membrane mean curvature $\nabla^2 h/2$, and is connected to the difference $C_1$ between the spontaneous curvatures of lipids A and B, through $\Lambda=-\kappa C_1$ (here $C_{\rm sp}=C_1 \phi$). A value $|\Lambda| \simeq 5$~pN has been proposed by fitting experimental data~\cite{Komura2006}. It is quite consistent with the typical values $\kappa \sim 10$~$k_{\rm B}T$ and $C_1 \sim 0.1$~nm$^{-1}$ discussed below. A value as large as $\Lambda \sim 100$~pN can reasonably be reached for more rigid Lo phases and stronger spontaneous curvatures.  

By integrating over the field $h(\br)$ (which in the present Gaussian approximation can be done~exactly), the effective Landau-Ginzburg Hamiltonian becomes 
\begin{equation}
\tilde H[\phi]= \int {\rm d}^2  \mathbf{r}\left[  w(\phi)+\frac{b'}2  (\nabla \phi)^2 + \frac{\Lambda^2 \kappa}{2 \sigma^2} (\nabla^2 \phi)^2\right]
\label{couplingLA2}
\end{equation}
where the stiffness $b$ has been ``renormalized'' to $b'=b-\Lambda^2/\sigma$, which can now be negative for either low enough surface tension $\sigma$, or low $b$ or low enough line tension $\lambda$, or high enough coupling $\Lambda$. It~signals the onset of a curvature instability of the membrane. If $b < \Lambda^2/\sigma$, a ``pattern'' spontaneously emerges with the characteristic length-scale (see also the beginning of Section~\ref{curv:comp:cp} below):
\begin{equation}
d \approx \sqrt{\frac{\Lambda^2 \kappa}{\sigma^2 | b'|}}.
\end{equation}

Note that this field-theoretic approach does not predict whether the emerging patterns look like roundish ``bubbles'' of an A-rich phase in a B-rich phase, or more elongated domains, or labyrinthine structures, or even alternate stripes~\cite{Seul1995}. We shall come back to this point below. 

Cluster phases are another instance of patterned structure extensively studied in soft condensed matter physics because they also appear in a wide variety of contexts~\cite{Stradner2004}. For example, if charged colloidal particles are in suspension in water, they feel a long-range mutual repulsion in low ionic force~solutions. If small polymers  are added to the solution, colloids also experience a short-range attraction usually named a depletion interaction~\cite{Gotzelmann1998} (in addition to the classical steric repulsion). Above a critical colloid concentration, small colloid clusters emerge in equilibrium, the typical size~of which is also set by the attraction-to-repulsion ratio and the concentration~\cite{Stradner2004}. The underlying mechanism is as follows: short-range attraction tends to favor phase separation, but long-range repulsion between domains limits their growth. The macro-phase separation where a single large colloid assembly would be formed is not reached because the long-range repulsion destabilizes too large clusters. The ideas that we will present below amount to extend this mechanism to protein assemblies. Whereas the preceding field-theoretic approach applies to both lipid and protein nanodomains because it is based on local concentrations, the cluster-phase scenario is more specific~to protein clustering because it deals with assemblies of individual objects in a continuous solvent (proteins in the 2D lipid mattress), as illustrated in Figure~\ref{two:fields}b~\cite{Destain2008}.

\subsection{Weak-Segregation Limit in the Vicinity of a Critical Point}
\label{aboveTm}

In this section, we assume that the lipid mixture is close to the miscibility critical point~\mbox{\cite{Chaikin,Onuki2002,Honerkamp2009}}, the so-called weak-segregation limit. The temperature $T$ is therefore close to (just above or just~below) the critical temperature $T_c$. In this limit, density fluctuations play an important role and one can construct a field theory using the Landau-Ginzburg formalism that includes weak density gradients and therefore wide interfaces between fluctuating lipid domains, which do not have a precise shape. On~the~contrary, in the strong segregation limit, where $T$ is well below the critical~temperature~$T_c$, the~two separated phases form well defined lipidic domains with sharp boundaries, and one can compute the free energy of chosen patterns, controlled by the interfaces between domains. This~situation will be considered in the Section~\ref{belowTm}.

\subsubsection{Curvature-Composition Coupling in Planar Membranes}
\label{curv:comp:cp}

Spontaneous curvature $C_{\rm sp}$ measures the more or less pronounced membrane tendency to bend spontaneously. Several mechanisms can lead to local spontaneous curvature $C_{\rm sp} \neq 0$ of lipid bilayers. The most evident one in cells is the marked difference of composition between the inner (cytoplasmic) and the outer (exoplasmic) leaflets of the plasma membrane~\cite{Alberts}. Some lipids have a global conical shape (to say it briefly, either ``big'' head-group and ``small'' tail(s), or ``small''  head-group and ``big'' tails) and locally concentrating them in one of the leaflets leads to non-vanishing spontaneous curvature~\cite{Mouritsen2005,Jarsch2016}. More precisely:
\begin{enumerate}
\item[(a)] The difference in lipid composition between both leaflets is important in cellular membranes, and it is maintained by the active cell metabolism. It can lead to bilayer spontaneous curvature if both leaflets conspire in this direction, because the bilayer curvature results form the difference~in the spontaneous curvature of the monolayers~\cite{Mouritsen2005}. The spontaneous curvatures of the main lipids found in plasma membranes are listed in~\cite{Zimmerberg2006} and they can be as large as $0.3$~nm$^{-1}$ for cholesterol or 1,2-Dioleoyl-sn-glycero-3-phosphoethanolamine (DOPE). 
 This is global on the whole membrane, but it can be accentuated locally due to the membrane lateral heterogeneity. For example, it has been shown on the basis of coarse-grained molecular dynamics simulations that mean curvatures of about $0.1$~nm$^{-1}$ can be attained in asymmetric membranes containing separated Lo and Ld phases on one leaflet and pure unsaturated  lipid on the other leaflet~\cite{perlmutter}. 
\item[(b)] The difference in the aqueous solution composition on the two sides of the membrane is maintained by the cell~\cite{Alberts}.   As explained by Lipowsky in 2013, a difference of solute concentrations, including ions and small molecules, generically leads to spontaneous curvature when they adsorb onto the membrane surface, for purely entropic causes~\cite{Lipowsky2013}. The membrane ``bends away from the exterior compartment if the concentration $c_{\rm ex}$ in this compartment exceeds the concentration $c_{\rm in}$ in the interior compartment''. For a single solute with different concentrations \mbox{across the membrane}, the spontaneous curvature is given by
\begin{equation}
C_{\rm sp} = \frac{k_{\rm B}T}{4\kappa} \,  \ell \, \Gamma_{\rm max} \, \frac{c_{\rm ex} - c_{\rm in}}{K_d},
\end{equation}
where  $\ell$ is the membrane thickness, $\Gamma_{\rm max}$ is the maximal surface density by adsorption and $K_d$ is the equilibrium constant of adsorption. Putting realistic numbers in this relation (in~particular $\kappa \approx 20 k_{\rm B}T$ and $\ell \approx 4$~nm~\cite{Phillips2015}), one gets $ C_{\rm sp} \approx 1/20$~nm$^{-1}$. The membrane itself is supposed~to~be up-down symmetric here. The adsorption of biopolymers is also examined~in this work~\cite{Lipowsky2013}. However, it is demonstrated to lead to smaller spontaneous curvature in realistic \mbox{regimes of parameters}. 
\item[(c)] The area difference between both leaflets can also lead to global spontaneous curvature. For~instance, an area difference of $\sim 10$\% leads a to spontaneous curvature  $C_{\rm sp}\sim 10^{-2}$~nm$^{-1}$~\cite{Zimmerberg2006}. This is the keystone of the area-difference-elasticity (ADE) model that has been developed to explain the rich shape variability of homogeneous lipid vesicles, in particular in function of their reduced volume $v$
~\cite{Mouritsen2005,Miao1994}. 
\end{enumerate}

Locally curving a membrane under tension leads to an increase of the Helfrich energy due~to the induced area excess. We shall see below that this energy penalty is lower when splitting a large curved domain into smaller ones and is then favorable, in spite of the increased interfacial energy penalty.
In~this respect, the field theory given above was very basic as compared to the actual developments~of the three last decades, an overview of which is now proposed. In this section, we~focus~on theoretical results obtained above the critical temperature where Gaussian approximations are useful. Presumably they remain valid in the vicinity of the critical point. Many results are similar~to what has been observed~in the context of micro-emulsions, made of a bicontinuous mixture of oil, water and amphiphile where one order parameter $\phi$ is now the local amphiphile concentration and the other one is the local concentration difference between oil and water~\cite{Roux1990,Gompper1994}.  

To be more quantitative, the membrane structure is well characterized by the structure factor (or power spectrum) $S(q)$, which is the Fourier transform of the spatial correlation function of lipid composition~\cite{Chaikin,Honerkamp2009}. At the level of the linear response, $S(q)$ characterizes the amplitude of the response~to an external perturbation of the local composition Fourier transform $\hat \phi(q)$ at a wave-vector~$q$. It is therefore the generalization of the susceptibility of the system at a given wave-vector~$q$.

At the Gaussian level where $w(\phi)=m\phi^2$ (with $m>0$) one can generically write the effective Landau-Ginzburg hamiltonian as
\be
\tilde H[\phi]=\frac12\int_0^\infty \frac{q{\rm d} q}{2\pi}S^{-1}(q) |\hat\phi(q)|^2
\ee

The classical structure factor for an inhomogeneous mixture has the form $S(q)=1/(\alpha(T-T_c)+bq^2)$, a Lorentzian of half-width $\xi_{\rm OZ}$. It diverges at the transition $T=T_c$ at $q=0$, the signature of a macrophase separation. By plugging in the coupling to the shape fluctuations, it is simple to show from \eq{couplingLA2}, that $S(q)$ has a second maximum at $q_c=\sqrt{|b'|\sigma^2/(\Lambda^2\kappa)}$ when $b'<0$. By introducing the Helfrich correlation length $\xi=\sqrt{\kappa/\sigma}$ and using $\Lambda=-\kappa C_1$, one can rewrite
\be
q_c^2=\xi^{-2}\left[1-\left(\frac{C_1^*}{C_1}\right)^2\right]\quad\mathrm{for}\quad C_1>C_1^*\equiv\frac{\sqrt{b\sigma}}{\kappa}.
\label{qc1}
\ee

Hence this  secondary maximum occurs as soon as $C_1$ is larger than $C_1^*$, i.e., for relatively large difference in spontaneous curvatures. Indeed, we recall that $C_1$ is the difference between the spontaneous curvatures of both phases, in other words $C_{\rm sp} (\phi) = C_0 + C_1 \phi$ at the lowest order in $\phi$.

This phase is a homogeneous (or liquid) but structured one, also called ``Structured Disordered'' (SD) phase,
 where correlations between lipids at a typical distance $1/q_c$ exist, which reflects a tendency towards order. The binary mixture is more susceptible to form transient structures of this size \mbox{when it is perturbed}. This regime of structured liquid on a length on the order of $\xi$ is consistent with the nanodomains of size ranging from 10 to 100~nm, for realistic cell membrane elastic parameter values~\cite{schick2012,shlomovitz}.

Within this picture, this secondary maximum diverges when $m=|b'|q_c^2/2$, i.e., for $T^*=T_c+|b'|q_c^2/(2\alpha)>T_c$~\cite{Brazovskii1975}.~Note that $T^*$ depends on $C_1$ through $q_c$. It is a monotonically increasing function of $C_1$. Hence a modulated phase appears for temperatures $T\leq T^*$  where the system undergoes mesophase separation~\cite{schick2012,shlomovitz,sunilkumar} which has been observed in simulations~\cite{perlmutter,stevens}. The~occurrence of stripe phases, square phases and hexagonal phases has been predicted using the single mode approximation and phase diagrams have been constructed for two and even three component fluids~\cite{sunilkumar}. 

Schick and coworkers argued that in plasma membranes both modulated phases (for which $S(q)$ diverges at $q_c$) and structured liquid (as soon as $C_1>C_1^*$)  can be observed~\cite{schick2012,shlomovitz,shlomovitz2014}. This would be the origin of lipid rafts, the typical size of which is $2\pi/q_c\simeq 100$~nm.  Note that this domain size is equal or larger than the Helfrich correlation length, $\xi$, according to \eq{qc1}, that is fixed by the balance between bending and surface energy. The classical interpretation for the criterion $C_1>C_1^*$ is the following: creating a bent domain is possible if the gain in bending energy $\approx \xi^2\kappa C_1^2$, is larger than the cost in demixing, $\approx b(\nabla\phi)^2 \xi^2 \approx b$. The last relation comes from the fact that on domains of size $\approx \xi$, $|\nabla\phi| \approx \xi^{-1}$. 

In references~\cite{mackintosh,Gueguen2014,Shimobayashi2016}, it is argued that a more correct way to introduce the coupling between composition and curvature is to write explicitly in the Helfrich Hamiltonian that the spontaneous \mbox{curvature $C_{\rm sp}$ is a} function of $\phi$. The full bending energy term in~\eq{couplingLA} is then replaced by $\frac{\kappa}2\int [\nabla^2 h (\br)-C_1\phi(\br) ]^2 {\rm d}^2 \br$ (if $C_0=0$ for sake of simplicity). This amounts to set $\Lambda = -\kappa C_1$~as already stated and more importantly to add a new composition-dependent surface tension term $\kappa C_1^2\phi^2(\br)/2$ in the Hamiltonian~\eq{couplingLA}. The inverse of the structure factor becomes $S^{-1}(q)=\alpha (T-T_c)+bq^2+\kappa C_1^2/[1+(q\xi)^2]$ which leads to an additional effective interaction mediated by the elastic membrane between lipids of the same kind,
\be
V(r)=\kappa C_1^2\int_0^\infty \frac{J_0(qr)}{1+(q\xi)^2}2\pi q\mathrm{d} q=2\pi\kappa C_1^2 K_0\left(\frac{r}{\xi}\right)
\label{VdeR}
\ee
where $J_0$ is the Bessel function of 0th order and $K_0$ is the modified one. Hence the interaction is repulsive due to higher membrane deformation (and therefore elastic cost) between similar lipids when they are closer. This interaction of bending nature has range $\xi$ and an amplitude set by $\kappa C_1^2$ and~is screened at large distances $r \gg \xi$ due to the surface tension cost.

It also modifies the critical temperature through $T'_c = T_c - \kappa C_1^2 / \alpha$, because the parameter $m$ in Equation~\eqref{w} becomes $m+\kappa C_1^2$, as well as  the value of $q_c$ according to~\cite{Gueguen2014}
\be
q_c^2=\xi^{-2}\left(\frac{C_1}{C_1^*}-1\right)
\label{qc2}
\ee
for $C_1>C_1^*$. Hence in this case, for $C_1>2C_1^*$, i.e., for large spontaneous curvature differences, the domain size can be smaller than $\xi$. More importantly, this new term forbids the formation~of mesophases by excluding the divergence of the structure factor at finite $q$, and only structured disordered phases can occur.

Using this curvature-induced mechanism, MacKintosh considered the alternative effect a quadratic confinement term $K h^2(\mathbf{r})$ in the Hamiltonian altogether with a vanishing surface tension. This also leads to modulated phases~\cite{mackintosh}. However, the characteristic wave vector is different since the term $\sigma q^2$ is replaced by a constant $K$ in the Fourier space. The second maximum of the structure factor then obeys $q_c^2\geq\kappa C_1^2/(2b)$. This confinement can also be seen as a model for the cell wall pinning by proteins {and/or tethering to the actin cortex~\cite{Wingreen2008,Agudo2017} (the membrane tension was restored in~\cite{Agudo2017}, and the analytical treatment was performed in spherical geometry there)}. 

The Leibler-Andelman mechanism described above does not properly take into account the bilayer structure of the membrane. It can be fully relevant when some biomolecules adsorb only onto the external leaflet as in the model system by Shimobayashi~et al.~\cite{Shimobayashi2016} where the ganglioside GM1~is~added~in the solution. However, although the average lipid composition of real \mbox{biomembranes is in} general asymmetric~\cite{shlomovitz,shlomovitz2014}, both cases where the local lipid compositions~of the two leaflets are correlated (registration) or anti-correlated (anti-registration) are in principle possible~\cite{Gueguen2014,Williamson2015}.

\begin{figure}[h]
\begin{center}
\includegraphics*[width=15cm]{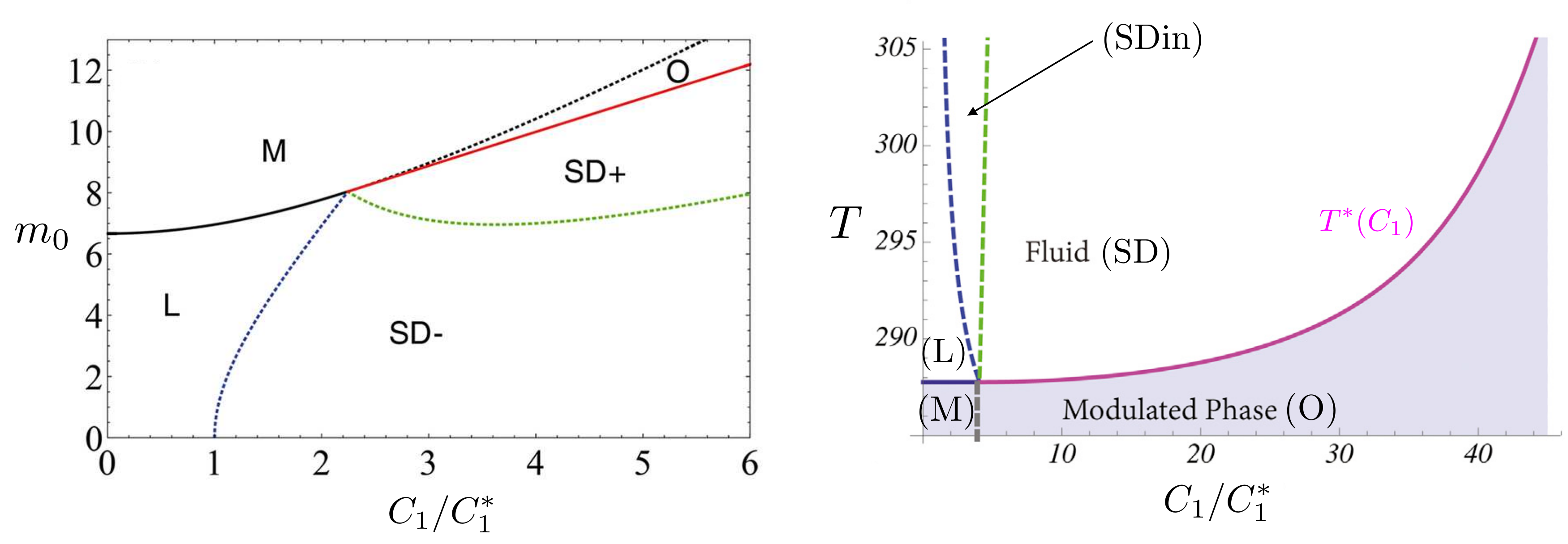} \\
\end{center}
\quad \hfill (\textbf{a}) \hfill \hspace{3cm} (\textbf{b}) \hfill \quad
\caption{Examples of phase diagrams showing the formation of curvature-induced domains in planar membranes. Each diagram shows 3 phases: the macrophase separation (M), the modulated phase (or mesophase) (O), and the liquid phase which can be simple (L) or structured disordered (SD) with transient domains. (\textbf{a})~Coupling constant between the two leaflets $m_0$ versus $C_1/C_1^*$ (see text) from the model by Gueguen et al. in which the two leaflets have the same composition but with different averaged area fractions. SD- (resp. SD+) corresponds to curved (resp. flat) transient domains.  Here the bending modulus $\kappa$ does not depend on the phase state. Reproduced from~\cite{Gueguen2014}, with permission of The European Physical Journal (EPJ), Copyright 2014.  
(\textbf{b})~Temperature versus $C_1/C_1^*$ from the model by Shlomovitz and Schick in which each leaflet contains a different mixture of two lipids. (SDin) corresponds to curved transient domains in the inner leaflet only, and~(SD) in both leaflets. Adapted from~\cite{shlomovitz}, with permission from Elsevier, Copyright 2013. }
\label{curvature_phase_diagram}
\end{figure}

Hence a more detailed description of the bilayer requires the introduction of two composition~fields, which correspond either directly to the compositions in one of the two types~of~lipids, relative to the average ones, $\bar\phi_{1,2}$ in the two monolayer (1) and (2), $\psi_{1,2}=\phi_{1,2}-\bar\phi_{1,2}$, or to the linear combination of them $\psi_{\pm}=(\psi_1\pm\psi_2)/2$. This idea has first been proposed by McKintosh~\cite{mackintosh}, and studied recently by Shlomovitz and Schick~\cite{shlomovitz} for two different lipid types~in the two leaflets and by Gueguen et al.~\cite{Gueguen2014} for two different average compositions of the same couple~of lipids in the two leaflets. The phase diagrams become richer since the numbers of ``masses'' in the field-theory, $m=\alpha(T-T_c)$, is no more one but three, associated with the terms in $m_+\psi_+^2$, $m_-\psi_-^2$, and~$m_0\psi_+\psi_-$~in~the Hamiltonian (assuming that the coefficient $b$ is the same in the two leaflets). \mbox{In both cases}, in addition to the modulated phase and the macrophase separation regions, the liquid phase (i.e., no order at long range) is now divided in 3 sub-regions: the true liquid one, and the two structured disordered regions corresponding to the two fields 
$\psi_{+}$ and $\psi_{-}$ (Figure~\ref{curvature_phase_diagram}a).

The interpretation is different in both cases. Shlomovitz and Schick fix some values to the species in each leaflet and using the Flory mean-field model, they construct a phase diagram in the $(C_1/C_1^*,T)$ plane, as displayed in Figure~\ref{curvature_phase_diagram}b. In particular, they study the case where both leaflets have different critical temperatures, that we denote by $T_c^{\rm in}$ and  $T_c^{\rm out}$. For example, in asymmetric lipid bilayers made of sphingomyelin, phosphatidylcholine in one leaflet, and phosphatidylserine and  phosphatidylethanolamine in the other one, $T_c^{\rm in} \approx 200$~K and 
$T_c^{\rm out} \approx 300$~K. Furthermore, they~assume that the coupling constant $|\Lambda|=\kappa |C_1|$ is on the order of 100~pN which implies that if the membrane bilayer is in the fluid phase, this phase is a micro-emulsion. They show that if $T>T_c^{\rm in}$~a structured liquid can appear on each leaflet: on the inner leaflet only for intermediate values of $C_1/C_1^*$ (region~SDin~in the figure), and on both leaflets for larger $C_1/C_1^*$ values. For even larger values, a~mesophase \mbox{(or modulated phase)} appears. In particular in region SDin, the two leaflets are decoupled such that the upper leaflet still behaves as a ordinary liquid. The rafts, however, are~identified by those authors as the transient curved domains that may appear in the region denoted SD in the figure, where~both leaflets are in the structured liquid regime.  

On the contrary, Gueguen et al.~\cite{Gueguen2014} construct a phase diagram at fixed temperature in the $(m_0,C_1/C_1^*)$ plane. In principle many parameters indeed depend on $T$ in an ill-defined way, notably~the masses but also the surface tension~\cite{Gueguen2017} and anticipating how the phase diagram behaves with $T$~is quite challenging. Since the same mixture of lipids is assumed to be present on each leaflet, the~critical temperature $T_c$ is the same. However, due to the difference in concentration, the demixing temperatures $T_d$ can be different. The parameter $m_0$ is related to these different average concentrations in the two~bilayers. According to the Flory theory, one has 
\begin{equation}
m_0 = k_{\rm B} T \left[ \frac1{\bar \phi_1 (1-\bar \phi_1)} - \frac1{\bar \phi_2 (1-\bar \phi_2)} \right].
\end{equation}

For large $C_1$ a structured liquid regime appears with transient curved domains \mbox{(phase denoted SD-)} and if $m_0$ is increased one observes the occurrence of a structured liquid with both~curved, and flat and thicker, transient domains (phase SD+). If Lo lipid rafts are assumed \mbox{to be flat}, the transient thick domains appearing in the SD+ region of the phase diagram are good candidates for being rafts.

Another mechanism has been proposed by Schmid and coworkers to explain the occurrence of small flat nanodomains~\cite{Schmid2017,Meinhardt2013,Brodbek2016}. In this model, a composition-curvature coupling similar to the one described above is introduced for each leaflet composing the bilayer. An elastic cost is introduced when the bilayer {has a width different from the equilibrium one}, similarly to the hydrophobic mismatch penalty close to proteins inserted in the membrane~\cite{Dan1993}. When the two leaflets bend towards the exterior {or the interior} of the membrane, this elastic cost forbids the formation of large domains leading~to small nanodomains instead of a macrophase separation. Schmid et al. argued that these domains are candidates for lipid rafts, due to their small size of a few nanometers, fixed by a balance between the bending modulus and the area compressibility~\cite{Schmid2017}. 

\subsubsection{Bending Modulus-Composition Coupling in Planar Membranes}
\label{bending}

Apart from a coupling between the spontaneous curvature and the composition, it has been proposed that for symmetric membranes a coupling through a composition-dependent bending~modulus, $\kappa(\phi)$, should occur. For instance Lo phases are thicker than Ld ones, which~leads to $\kappa_{\rm Lo}>\kappa_{\rm Ld}$ as already discussed. The simplest way to deal with this fact is to interpolate linearly between both values as $\kappa(\phi)\simeq\kappa_0+\kappa_1(\phi-\phi_0)$. Hence a new term $(\kappa_1/2) \phi(\nabla^2 h)^2$ enters the Hamiltonian of the field theory, beyond the Gaussian order. In addition to the above coupling in $\phi(\nabla^2 h)^2$, additional terms can be introduced in $\phi(\nabla h)^2$ or higher-order ones where $\phi$ is replaced by $\phi^2$ or $(\nabla \phi)^2$.

On the other hand several works have studied the interaction between two rigid \mbox{inclusions in a} fluctuating membrane~\cite{Dan1993,goulian,Netz1995,Netz1997} and shown that it is attractive and proportional to $k_BT$ which means \mbox{that~it is a} membrane fluctuation-mediated interaction (see the so-called ``Casimir'' forces~in Section~\ref{cl:ph}). It is natural to anticipate that the more rigid Lo domains will also interact in a similar way. By performing a cumulant expansion, Dean and Manghi~\cite{dean} indeed showed that the resulting fluctuation-induced interaction between lipids is a Casimir-like attractive interaction $V(r) \propto-k_BT/r^4$~at intermediate distances. However, the interaction at short distances $V(r) \propto-k_BT\kappa_1w'(\phi_0)/(\kappa_0^2r^2)$, where $w(\phi)$ has been defined in Equation~\eqref{w}, is larger than the van der Waals attractive interactions and can be either repulsive or attractive depending on the sign of $\kappa_1 w'(\phi_0)$. In the case where $\kappa_1 w'(\phi_0)<0$, mesophases can occur due to this repulsive fluctuation-induced interaction for large values of $\kappa_1$ and small values of $w'(\phi_0)$. Note that it has been proposed using coarse-grained simulations that $\kappa(\phi)$, where $\phi$ is the area fraction of short amphiphiles, is not linear, as~assumed above. It decreases when $\phi$ increases for $\phi<0.6$ but increases slowly for higher values of $\phi$~\cite{Imparato2005,Brannigan2005}. These numerical works show that $-4 \leq \kappa_1 / (k_{\rm B} T) \leq2$ which suggests that the condition $\kappa_1 w'(\phi_0)<0$ can be reached in real systems.

\subsubsection{Vesicles}
\label{vesi}

We begin with an homogeneous vesicle.  All the above theories where developed for planar membranes for which the average curvature vanishes, $\int {\rm d}^2  \mathbf{r} \langle \nabla^2 h\rangle=0$, a constraint that is relaxed~in spherical geometry.
It is important to note that the vesicle can be studied either at constant \mbox{volume $V$ or at} constant pressure jump between the interior and the exterior of the vesicle, $\Delta p=p_{\rm in}-p_{\rm out}$. This pressure jump can be an osmotic one due to solutes to which the membrane~is impermeable, or from a theoretical perspective, a Lagrange multiplier to enforce the volume constraint. This pressure jump fixes therefore the projected area $A_s=4\pi R^2$ of the sphere having the same~volume, related to $R$ by $V=4\pi R^3/3$, and plays exactly the same role as the frame tension for planar membranes~\cite{Gueguen2017}. At the Gaussian order and for large surface tensions, $\sigma\gg \kappa/R^2$, both are connected through the Laplace law 
\be
\Delta p= 2 \sigma/R.
\ee 

In particular, this expression can be interpreted as follows: the limit $\sigma \rightarrow \infty$, where thermal fluctuations are suppressed, is equivalent to the high pressure limit $\Delta p\rightarrow \infty$, itself equivalent to the limit where the reduced volume $v \equiv 6 \sqrt{\pi}V/A^{3/2}$ goes to 1.

In contrast, for low surface tensions, $\sigma\lesssim \kappa/R^2$, the vesicle is no more quasi-spherical and a local equilibrium shape condition has been derived by Zhong-can and Helfrich~\cite{Zhong-can1989} 
\be
\Delta p= 2 \sigma H-\kappa[(2H)^2-C_{\rm sp}^2)]H-2\kappa \nabla^2H
\ee
which generalizes the Laplace law. Note however, that in the limit $\sigma\lesssim \kappa/R^2$, high order terms must be taken into account in the Hamiltonian, which renormalize the surface tension~\cite{Gueguen2017}.

\begin{figure}
\begin{center}
\includegraphics*[width=15cm]{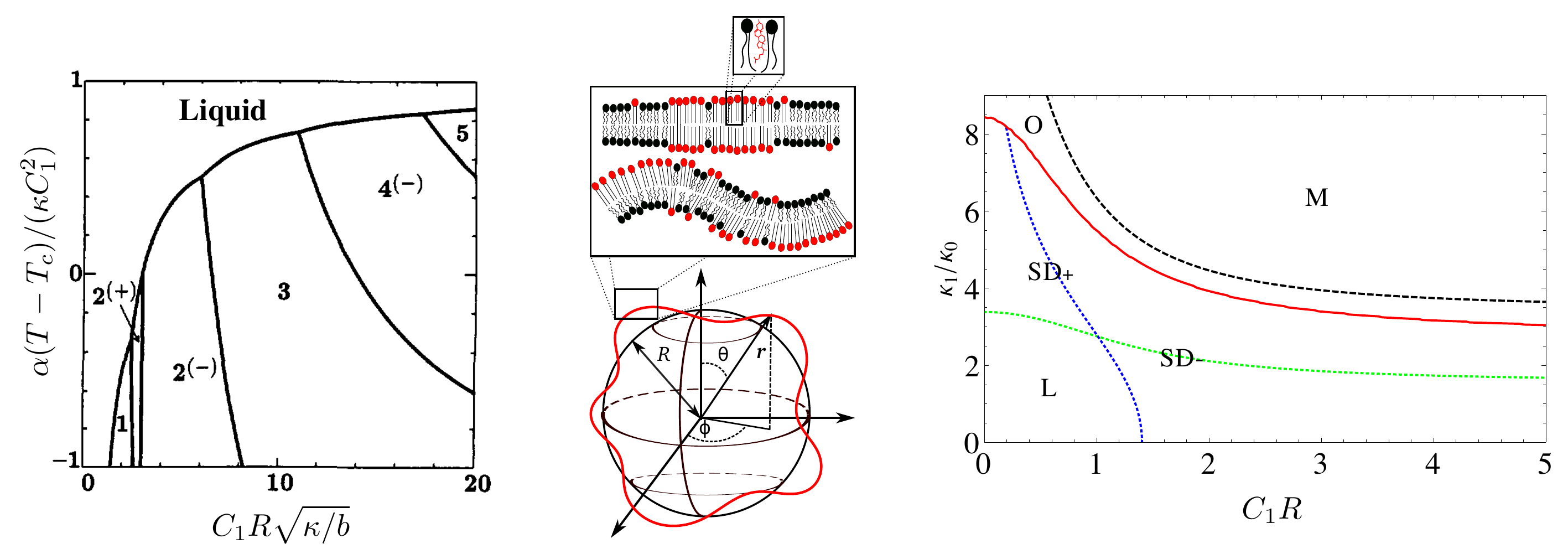} \\
\end{center}
\quad \hfill (\textbf{a})  \hspace{4cm}  (\textbf{b}) \hspace{6cm} (\textbf{c}) \hfill \quad
\caption{Examples of phase diagram showing the formation of curvature induced domains in vesicles. (\textbf{a})~AT high $T$ or low $C_1$, the liquid phase is homogeneous;  numbered regions correspond to modulated phases with the number of the most stable $\ell$-mode ($\ell=1$ to 5 here; $\ell=1$ corresponds to the macrophase separation). The superscript corresponds to the sign of $C_1$. The bending modulus $\kappa$ does not depend on the order parameter $\phi$ in the model. In this case, $C_0=\Delta p=0$. Adapted from~\cite{Taniguchi1994}. (\textbf{b})~Sketch of a quasi-spherical vesicle with two different types of lipids inducing either thicker (with a larger bending modulus $\kappa_0+\kappa_1$) or curved patches with a local spontaneous curvature $C_0+C_1$. (\textbf{c})~Associated phase diagram in the $(C_1R,\kappa_1/\kappa_0)$ plane. The symbols have the same signification as in Figure~\ref{curvature_phase_diagram}a.   (\textbf{b}) and (\textbf{c}) are reproduced from~\cite{Gueguen2014}, with permission of The European Physical Journal (EPJ), Copyright 2014.}
\label{vesicle_phase_diagram} 
\end{figure}

The field-theoretic framework presented above can be easily extended to the case of quasi-spherical vesicles made of a lipid mixture. The membrane fluctuations of single-component vesicles have been described by the Helfrich hamiltonian expressed in the spherical  harmonics basis $(Y_{\ell}^m)$~\cite{Helfrich1986,Milner1987,Seifert1997,Barbetta2010}. Taniguchi et al.~\cite{Taniguchi1994} investigated the equilibrium shapes of two-component vesicles using the same coupling as in \eq{couplingLA} when the temperature $T$ is just below $T_c$ \mbox{(weak segregation limit)}, and at fixed pressure jump $\Delta p$. By performing a linear stability analysis and using the single-mode approximation, they obtained phase diagrams focusing on the role played~by the domain boundary energy (parameter $b$) which favors fewer domains and the coupling energy  proportional to $\Lambda$ which favors high-$\ell$ mode states. They found a large number of domains for $\ell=3,4,5...$ (the size of which is on the order of the vesicle one) for very small values of $b$, even for $T>T_c$ (see Figure~\ref{vesicle_phase_diagram}a). This mechanism is exactly the same as the Leibler-Andelman one described in Section \ref{curv:comp:cp} for planar membranes.

The case of symmetric bilayer with $\kappa$ depending on $\phi$ has been studied numerically by \mbox{Feigenson et al}. for $T<T_c$ (see Section~\ref{carapace} below). In an attempt to unify both \mbox{types of approaches}, Gueguen et al. studied on an equal footing the cases where both the spontaneous curvature $C_{\rm sp}$ and the bending modulus $\kappa$ depend on the composition $\phi$~\cite{Gueguen2014}, for quasi-spherical vesicles~at fixed surface tension $\sigma$. When the vesicle is made of two types of lipids of different sizes, $C_{\rm sp}$~increases when the composition is locally non-symmetric between the two leaflets whereas $\kappa$ increases when the longer lipids are in register (see Figure~\ref{vesicle_phase_diagram}b). By again choosing a linear interpolation, $C_{\rm sp}=C_0+C_1(\psi_{\rm out}-\psi_{\rm int})/2$ and $\kappa=\kappa_0+\kappa_1(\psi_{\rm out}+\psi_{\rm int})/2$, phases diagrams in the plane $(C_1,\kappa_1)$ are constructed at Gaussian order (see Figure~\ref{vesicle_phase_diagram}c). Even for $C_1=0$ (no curvature-induced mechanism), a structured liquid phase occurs (SD+) with transient highly correlated flat raft-like domains of larger bending rigidity, whereas no transient curved domains appear. They 
appear for high enough values of $\kappa_1>\kappa_1^*$, with  
\be
\kappa_1^*= \frac{2 \sqrt{2 b \kappa_0}}{(2-C_0R)\sqrt{6-C_0R}} \, \frac{6 + \sigma R^2 /\kappa_0 - C_0R(2-C_0R)}{\left[ \sigma R^2 (2+C_0R) /\kappa_0 - 6(2-C_0R)+(C_0R)^3\right]^{3/2}}
\label{kappa1star}
\ee
if $m_0=0$. If $m_0 \neq 0$, a more complex formula is given in reference~\cite{Gueguen2014} (see Figure~\ref{vesicle_phase_diagram}c). Again, the structured disordered phase occurs when the elastic cost associated with difference $\kappa_1$ in bending moduli is larger than the interfacial energy~$\propto b$. 

This regime is experimentally accessible given the measured values of $\kappa_{\rm Lo}/ \kappa_{\rm Ld}$ already discussed. This effect cannot be seen in planar membranes. Indeed, $\kappa_1^*$ is finite provided that $C_0$ is different from the average curvature $2/R$ imposed by the vesicle volume, whereas $C_0R=2$ is imposed in the large $R$ limit corresponding to the planar case~\cite{Gueguen2014}. For non-vanishing $C_1$, a large region with structured phases emerges in the phase diagram, as soon as $C_1\gtrsim C_0$ and $\kappa_1\gtrsim 3\kappa_0$ (see Figure~\ref{vesicle_phase_diagram}c).\

\subsection{Strong Segregation Limit}
\label{belowTm}

Well below the demixing temperature, appealing to field-theoretic techniques is more involved because it is necessary to take into account higher order $\phi^4$ terms in the Hamiltonian. Alternative~effective theories have been proposed in the literature, the most popular of which are now~presented. They use the fact that in the strong segregation limit, the large line tension prevents the boundary $\mathcal{B}_{\rm Lo-Ld}$ between both phases to fluctuate much at short wavelengths. The phase-boundary energy can be written as its length $|\mathcal{B}_{\rm Lo-Ld}|$, which is well defined, times the line tension, as we already explained it in the introduction of this Section~\ref{equil}~\cite{Seul1995}:
\begin{equation}
\frac{b}2 \int_S {\rm d}^2 \mathbf{r} (\nabla \phi)^2 = \lambda | \mathcal{B}_{\rm Lo-Ld} |  \quad {\rm with} \quad \lambda \propto  b\Delta\phi^2/\xi_{\rm OZ}.
\label{line:tension:def}
\end{equation}

In the strong segregation limit, the typical values of $\lambda$ are measured to be on the order of a few~pN~\cite{Garcia2007,Usery2017,Baumgart2003,Semrau2008,Ursell2009}.

This idea was first fully exploited in 1992 by Andelman and co-authors~\cite{Andelman1992,Kawakatsu1993}, in the case where the composition field $\phi$ and the curvature $\nabla^2 h$ are coupled as in Equation~\eqref{couplingLA}, through the coupling constant $\Lambda$. In simplified geometries, i.e., striped planar membranes, and cylindrical or axially-symmetric vesicles, they could show that modulated mesophases are more stable than a macrophase separation in some favorable circumstances. As compared to the condition $b < \Lambda^2/\sigma$ for mesophase separation discussed below~Equation~\eqref{couplingLA2}, being far from the critical point leads to a different discussion. 
In the planar case, the emerging wavelength is $d^* \propto \xi (\sqrt{  \sigma \, \kappa \, b \, |m|} / \Lambda^2)^{1/3}$  when $d^* \ll \xi$, i.e., in the low surface tension limit. In the opposite limit, $d^*$ diverges logarithmically when the membrane tension $\sigma$ approaches the limiting tension 
\begin{equation}
\sigma_{\rm cr} \equiv \Lambda^4/(16 \, \kappa \, b \, |m|),
\label{sigma:lim}
\end{equation}
signaling the onset of macro-phase separation. 

In the particular vesicle geometries tackled in this preliminary work, the role of the pressure difference $\Delta p$ across the membrane was explored. At fixed membrane area, if $\Delta p=0$, then the volume~is free to fluctuate and the vesicle adopts the most favorable geometry where mono-phasic domains are spherical caps with their prefered curvature. If $\Delta p \neq 0$, numerical integration of the ordinary differential equation 
governing the vesicle shape shows how the vesicle is deformed by pressure. When the interior pressure is larger than the exterior one ($\Delta p > 0$ with our notations), the number of domains grows with $\Delta p$ in spite of the growing interfacial energy in Equation~\eqref{line:tension:def} because this allows the enclosed volume $V$ to be larger as compared to a macrophase separation. 

A rich literature has followed this pioneering work to account for mesophase separation in the context of coexisting Lo and Ld phase separated by a marked boundary. 

\subsubsection{Domain Buckling Induced by Line Tension---Spontaneous Symmetry Breaking}
\label{buckling}

\begin{figure}[h]
\begin{center}
\ \includegraphics*[width=14cm]{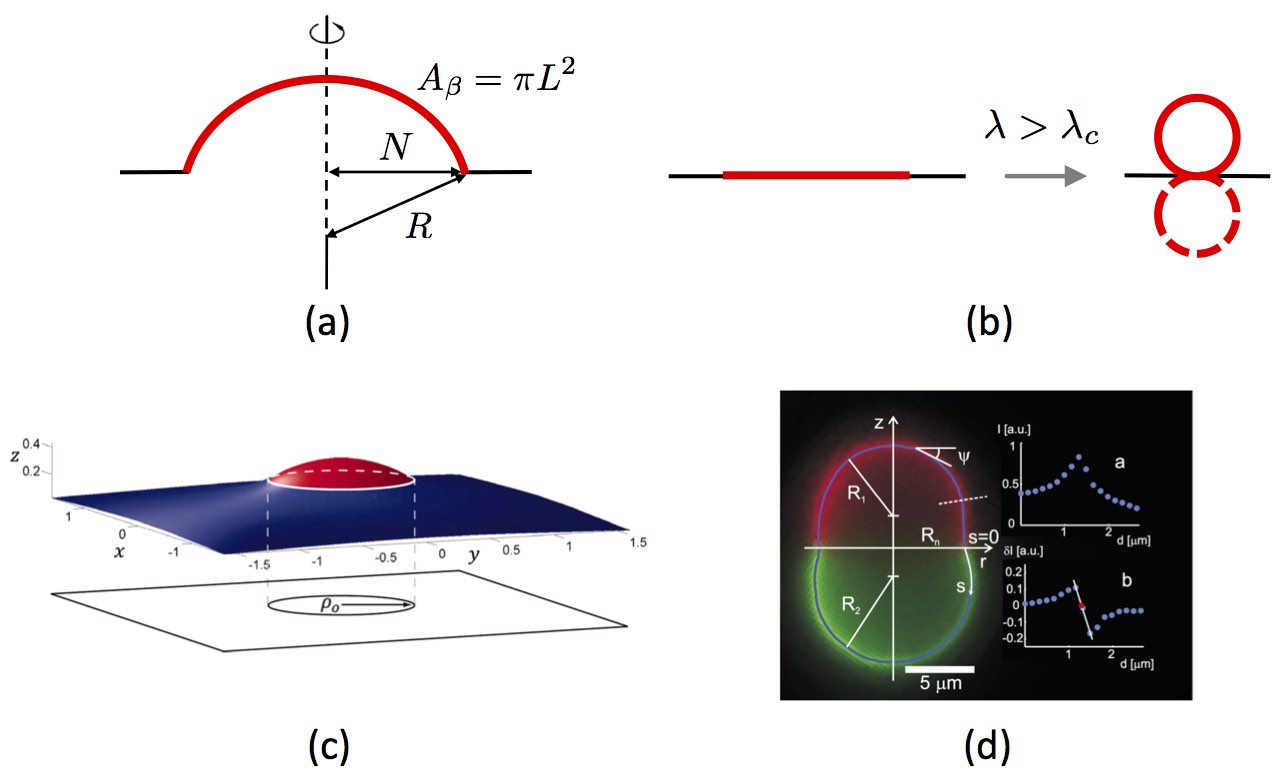} 
\end{center}
\caption{(\textbf{a}) In the case of plane tensionless membranes, a first approximation consists of considering spherical caps of Lo phase (or $\beta$ phase, in red) of area $A_\beta = \pi L^2$ in an otherwise infinite Ld membrane~\cite{Lipowsky1992}; the spherical cap radius is denoted by $R$ and the radius of the circle delimiting the phase boundary is $N$. (\textbf{b}) When the line tension $\lambda$ exceeds a critical value $\lambda_c$ (Equation~\eqref{lambdac}), the previously planar domain buckles to a complete sphere in the tensionless case. Owing to the spontaneous symmetry breaking principle, the domain is equally likely to buckle upward or downward (up-down symmetry). (\textbf{c}) In a taut membrane with surface tension $\sigma$ budding is incomplete. Reproduced from~\cite{Ursell2009} with permission from the Proceedings of the National Academy of Sciences USA, Copyright 2009. (\textbf{d}) In the case of giant unilamellar vesicles (GUV), fluorescence microscopy experiments on ternary mixtures of 1,2-Dioleoyl-sn-glycero-3-phosphocholine (DOPC)/Sphingomyelin (SM)/cholesterol below the demixing temperature distinguishes the Lo and Ld phases appearing in red and green respectively. Mathematical modeling then enables the extraction of physical parameters such as bending moduli $\kappa$ and line tension $\lambda$ at the Lo/Ld interface. Reproduced from~\cite{Semrau2008}, with permission from the American Physical Society, Copyright 2008.} 
\label{lipowsky:Fig}
\end{figure}

As illustrated in Figure~\ref{lipowsky:Fig}, the first mechanism that we present is based upon spontaneous up-down symmetry breaking induced by the strong line tension $\lambda$. We first focus on planar membranes before dealing with closed vesicles.

\begin{enumerate}
\item Planar membranes---We begin
 with the simplest form of this mechanism, as proposed in planar geometry by Lipowsky in 1992~\cite{Lipowsky1992}. We consider a single membrane Lo domain (denoted~$\beta$-phase in this work) in a large planar Ld membrane ($\alpha$-phase). Well below the demixing~temperature, the boundary shape is close to a circle to minimize the interfacial~energy. The total domain area is denoted by $A_{\beta} \equiv \pi L^2$ ($L$ is its radius in the membrane plane). Lipowsky~first assumes that the surface tension is vanishingly small ($\sigma=0$).  If it buds in the third dimension, the domain adopts the shape of a spherical cap supported by a sphere~of radius $R$, while the surrounding membrane remains flat (Figure~\ref{lipowsky:Fig}a). The interface is now~a circle~of radius $N\leq L$. Adopting a mechanical approach where fluctuations are ignored, the~total elastic energy $E_{\rm el}$~of the domain is given by the sum of two antagonist contributions: the boundary line-energy $E_{\rm bound} = 2 \pi N \lambda$ that is proportional to the domain boundary length and tends to minimize it (by protruding in the third dimension); and the elastic Helfrich energy $H_{\rm Helfrich}$ which disfavors~bending. For a fixed domain area $A_{\beta}$ the value of the cap radius $R$ is obtained by minimizing $E_{\rm el}(R)$. A natural length-scale $\xi_I=\kappa/\lambda$ can be introduced, called the ``invagination length''. If $\kappa_\beta \sim 100 k_{\rm B} T$ for the Lo phase~\cite{Semrau2008} and $\lambda \approx 1$~pN far from the critical point, then~$\xi_I \approx 400$~nm. When getting closer to the critical point, $\lambda$ decreases as $(T_c-T)^\nu$ with $\nu$ a universal critical exponent equal to 1 in 2D biphasic systems in the 2D Ising universality class~\cite{Chaikin,Lipowsky1992} and $\xi_I$ grows. We shall come back to these values later in the Discussion Section.

\quad\quad  Lipowsky shows that  if $L<4\xi_I$, then the optimal geometry is a flat domain ($N=L$); conversely, if $L>4\xi_I$, it is a complete sphere ($N=0$), protruding upward or downward with equal probability. Differently said, this simple model without surface tension proposes that above a critical line tension 
\begin{equation}
\lambda_c = 4 \frac{\kappa_\beta}{L},
\label{lambdac}
\end{equation}
spontaneously breaking the up-down symmetry is energetically favorable because it reduces~to zero the interfacial energy cost in spite of the increased bending energy (Figure~\ref{lipowsky:Fig}b). The~transition~is first order. When $C_{\rm sp}  \neq 0$, the symmetry between the two sides of the membrane~is explicitly broken. If the value of $L/\xi_I$ is sufficiently small, the minimization~of energy $E_{\rm el}(R)$ predicts intermediate equilibrium values of $N$, between 0 and $L$. This situation for which the bending energy and the interfacial energy balance each other is called incomplete budding, or ``dimpling''.

\item Additional role of surface tension---The case where the surface tension is finite, $\sigma>0$, has been explored in detail in~\cite{Ursell2009}. As bending stiffness, membrane tension applied in the membrane plane favors flat domains and comes in opposition to interfacial energy minimization. In this case~also, and without necessarily appealing to spontaneous curvature, incomplete budding occurs above a critical line tension, through the spontaneous symmetry breaking principle (Figure~\ref{lipowsky:Fig}c). The transition from flat to dimpled domains is now continuous whereas it was discontinuous without tension. More quantitatively, it is proven in this work that the critical line tension is given by $\lambda_c \simeq 8 \kappa_\beta /L$ in the limiting case where the domain area $A_{\beta} = \pi L^2 \ll \kappa_\beta/\sigma$. Here $\kappa_\beta$ is the domain stiffness, which can be different from the surrounding membrane one, $\kappa_\alpha$. Coming back to the notations used in the  paragraph just above, this condition reads $L = 8\xi_I$ at the critical point, which is twice the transition value found when $\sigma = 0$. This means that in the interval $4 \xi_I \leq L \leq 8 \xi_I$, budding is energetically favored when $\sigma = 0$, but becomes less stable than the flat geometry as soon as $\sigma$ is positive, even if small. 

\quad\quad Furthermore, just above this critical value, the contact angle $\epsilon$ at the domain boundary scales as $|\epsilon| \propto \sqrt{\lambda/\lambda_c - 1}$. The Lo domain continuously but rapidly deviates from the flat state. By up-down symmetry, the domain is equally likely to bud upward ($\epsilon > 0$) or downward ($\epsilon < 0$).  When $C_{\rm sp}  \neq 0$, this symmetry is again explicitly broken, and phase diagrams can also be inferred~\cite{Ursell2009}. 

\item Vesicles---J\"ulicher and Lipowsky addressed the same question in the case of biphasic vesicles with spherical topology~\cite{julicher}. As above, different situations exist, but the up-down symmetry (more~precisely the exterior/interior symmetry in this case) is explicitly broken on a vesicle. As~stressed in the field-theoretic approaches presented in Section~\ref{vesi}, a new ingredient can come into play here, namely the conservation of the volume $V$ enclosed by the vesicle, or equivalently the pressure jump across the membrane, $\Delta p$, which can be controlled through the osmotic pressure difference. The control parameter is, e.g.,  the reduced volume $v \equiv 6 \sqrt{\pi}V/A^{3/2} \leq 1$, measuring the deviation to a sphere (for which $v=1$).  If two domains coexist as above, describing the membrane through an elastic continuum theory, the minimization of the total energy provides the so-called ``shape equations'', from which the equilibrium vesicle shape under the relevant constraints is derived. In particular, it depends on the relative area fractions and on the different parameters (bending moduli $\kappa$, saddle-splay moduli $\kappa_G$, spontaneous curvature~$C_{\rm sp}$), which can~in principle be different in the two phases. Indeed, even though it is not a pre-requisite, the difference between the bending moduli of the two phases now likely plays a role, contrary to the planar case, because both phases are bent in this geometry. 

\quad\quad  A rich phase diagram can be computed by minimizing the membrane energy, still neglecting thermal fluctuations. In this case as well, budding can be incomplete or complete, a closed vesicle then being connected to the main vesicle through an infinitesimal ``neck''. \mbox{However, a strong} volume constraint $v \simeq 1$ (or equivalently large $\sigma$), where the shape is quasi-spherical, can act against the budding process but does not, in general, suppress it. The reader can refer to~\cite{julicher} for further details. These results have been confirmed by numerical coarse-grained modeling (4-bead lipids and explicit solvent) based on dissipative particle dynamics, where both area and volume are conserved~\cite{laradji2004}. 

\item Experiments---Fluorescence microscopy experiments~\cite{Baumgart2003,Semrau2008,Veatch2002} have later validated this theoretical approach on free-floating giant unilamellar vesicles (GUV) made of ternary mixtures~of saturated lipids, unsaturated lipids and cholesterol, well below the demixing temperature, which~display separated Lo and Ld phases (Figure~\ref{lipowsky:Fig}d). In reference~\cite{Yanagisawa2008}, the reduced volume~$v$~of GUVs made of a DPPC
/DOPC/cholesterol mixture is controlled by varying the osmotic~pressure. If one starts from a spherical vesicle, domains bud (inward or outward according to the experimental conditions) when the enclosed volume decreases. 
Following these original~studies, a series of papers studied  the experimental counterpart of these theoretically predicted circular, budded Lo domains and established phase diagrams~\cite{Semrau2009,Ursell2009,Yanagisawa2007}. When the cholesterol concentration was increased above $\approx 35$\%, a reversal phenomenon was observed, now with Ld domains in a Lo continuous background. The domain sizes were typically observed to be in the micron scale. We have previously explained that if $\sigma \neq 0$~\cite{Baumgart2003}, then the critical radius $L$ above which domains buckles is $8 \xi_I \sim$ few~$\mu$m with the above value of $\xi_I \approx 400$~nm. Experiments and theory are compatible. Even though in a less evident manner, AFM experiments also suggest that budding exists in planar geometry~\cite{Garcia2007}, as predicted by theoretical approaches in the relevant regimes of parameters. 

\item Elastic interaction between budded domains---In these experiments, it is also observed that domains sometimes coalesce~\cite{Baumgart2003,Yanagisawa2007} but that this process is very slow and does not follow the usual laws of coarsening~\cite{Bray1994}. The reason is that budded domains repel each other when they come in close proximity because they deform the elastic membrane, in an enhanced way if they are very~close. This repulsion has even be very well quantified experimentally~\cite{Semrau2009,Ursell2009,Yanagisawa2007} and shown to be compatible with theoretical predictions. A supposedly metastable configuration is then observed with long but finite lifetime. After several hours, all Lo domains eventually coalesce and one ends with a complete macro-phase separation. Note also that coarsening is not always trapped and that the existence of normal coarsening has been correlated to a vesicle reduced volume $v$ very close to 1~\cite{Yanagisawa2007}. Indeed, budding requires excess area that is only available if the vesicle is at least slightly deflated. 

\quad \quad As a matter of fact, the complete proposed scenario is as follows: after quenching below the demixing temperature and once domain have nucleated, normal \mbox{coarsening is initiated}, with~small but growing nanoscopic domains. Being small, these domains are flat as demonstrated above~\cite{Semrau2009}. When their size reaches the critical value, all these domains suddenly buckle and coarsening is then trapped in the metastable state~\cite{Ursell2009,Yanagisawa2007}. The lateral organization~of domains observed on phase-separated  Sphingomyelin(SM)/DOPC/cholesterol 
vesicles in~\cite{Rozovsky2005} has been attributed~to this inter-domain repulsion, and the force between domain measured. Strong~slowing-down of domain coarsening observed in DPPC/DOPC/cholesterol GUVs~\cite{Saeki2006} was also \mbox{attributed to budding}, even though the inter-bud repulsion was not explicitly \mbox{appealed to in} this work. In contrast, when budding is avoided on sufficiently taut~vesicles, no~slowing-down is observed with respect to the expected dynamical exponent~\cite{Stanich2013}. 
\end{enumerate}

We have presented here the membrane buckling mechanism due to the high line tension $\lambda$~at the domain boundary well below the phase-transition temperature. It does not require any spontaneous curvature $C_{\rm sp}$ of the Lo phase nor any difference of bending rigidities $\kappa$ between both~phases. However,~cell~plasma membranes are believed to be relatively close to the phase transition temperature, for which the value of $\lambda$ is lower than the pN order of magnitude considered above, and the lower limit of the domain radius $L$ required to have buckling is thus significantly larger than 1~$\mu$m. \mbox{Below~this~radius}, domains coarsen and $L$ continuously grows with time. It is thus unlikely that the main driving mechanism leading to the observed nanodomains in cells is buckling. We shall come back to this issue in the discussion section. 

\subsubsection{Competing Interactions: Phase-Dependent Bending Modulus}
\label{carapace}

Bending elastic moduli of lipid bilayers $\kappa$ are known to depend on the lipid species and on the phase state (Lo or Ld) as already discussed~\cite{rawicz}. When peptides or proteins are included in the bilayer, the effective value of $\kappa$ also significantly depends on their concentration~\cite{Dimova2014,Fowler2016,Usery2018}. In the vesicle~geometry, the bending elastic energy necessarily plays a role since the membrane is always globally curved. The way the different membrane constituents spatially organize in the membrane plane can be more or less favorable with respect to this bending constraint.

\begin{figure}[h]
\begin{center}
\begin{tabular}{cc}
\parbox{7cm}{\begin{tabular}{c}
\includegraphics*[width=6cm]{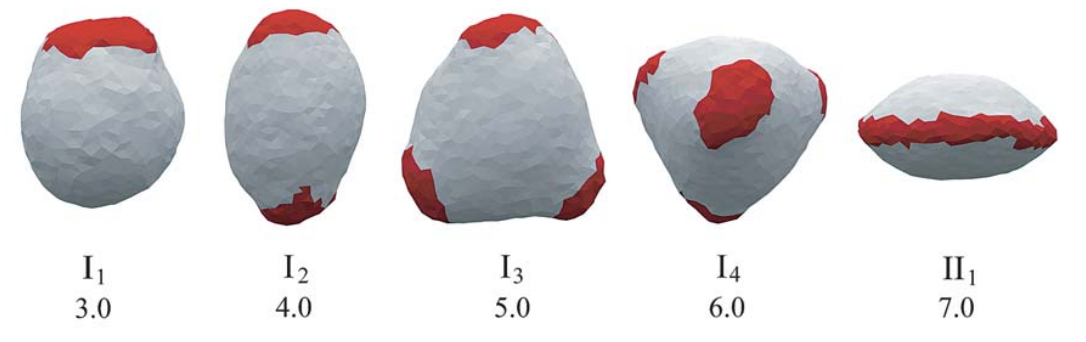} \\ (\textbf{a}) \vspace{0.5cm} \\ \includegraphics*[width=6cm]{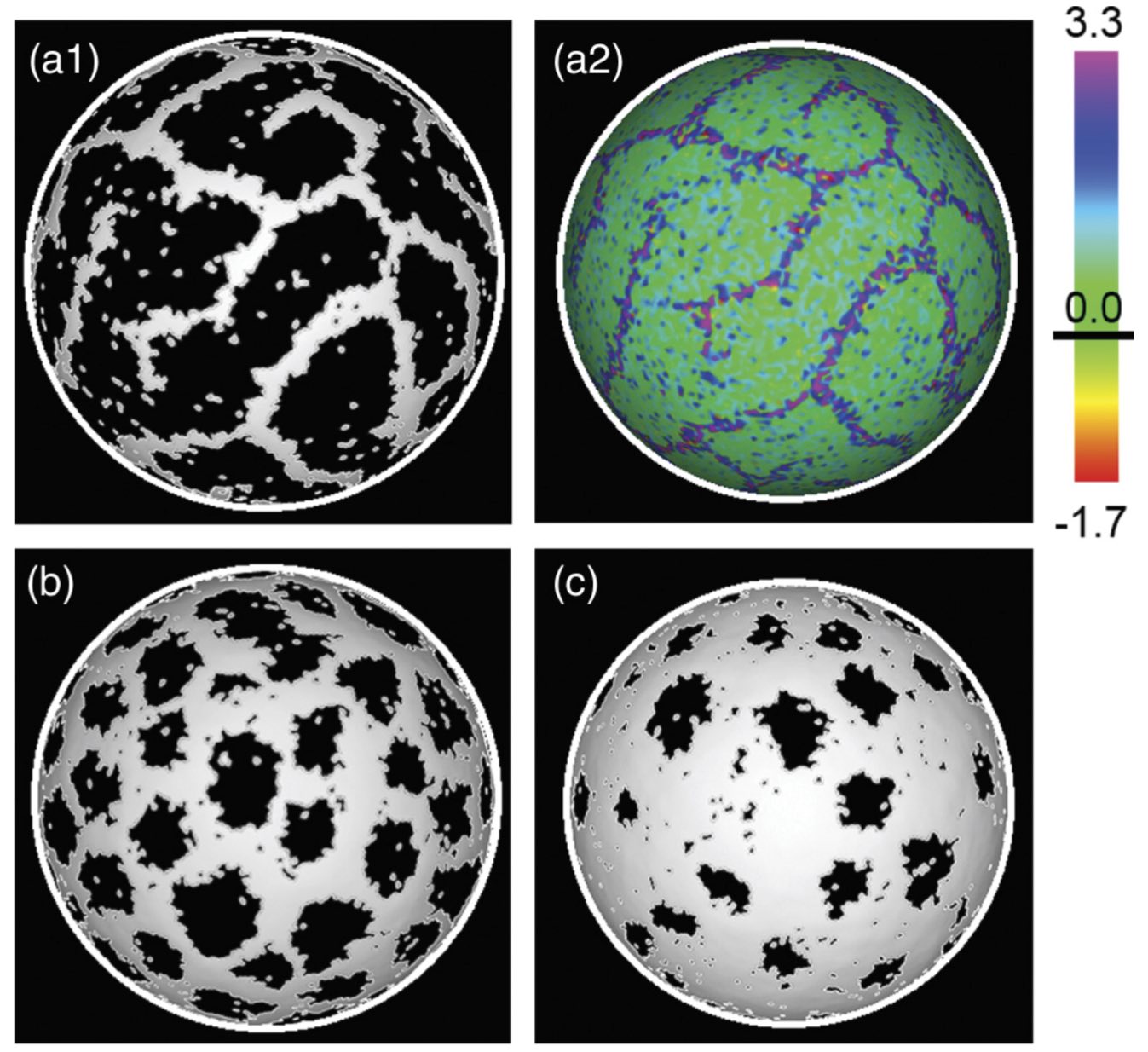} \\ (\textbf{b})
\end{tabular}}
& \quad \parbox{8cm}{\includegraphics*[width=7.5cm]{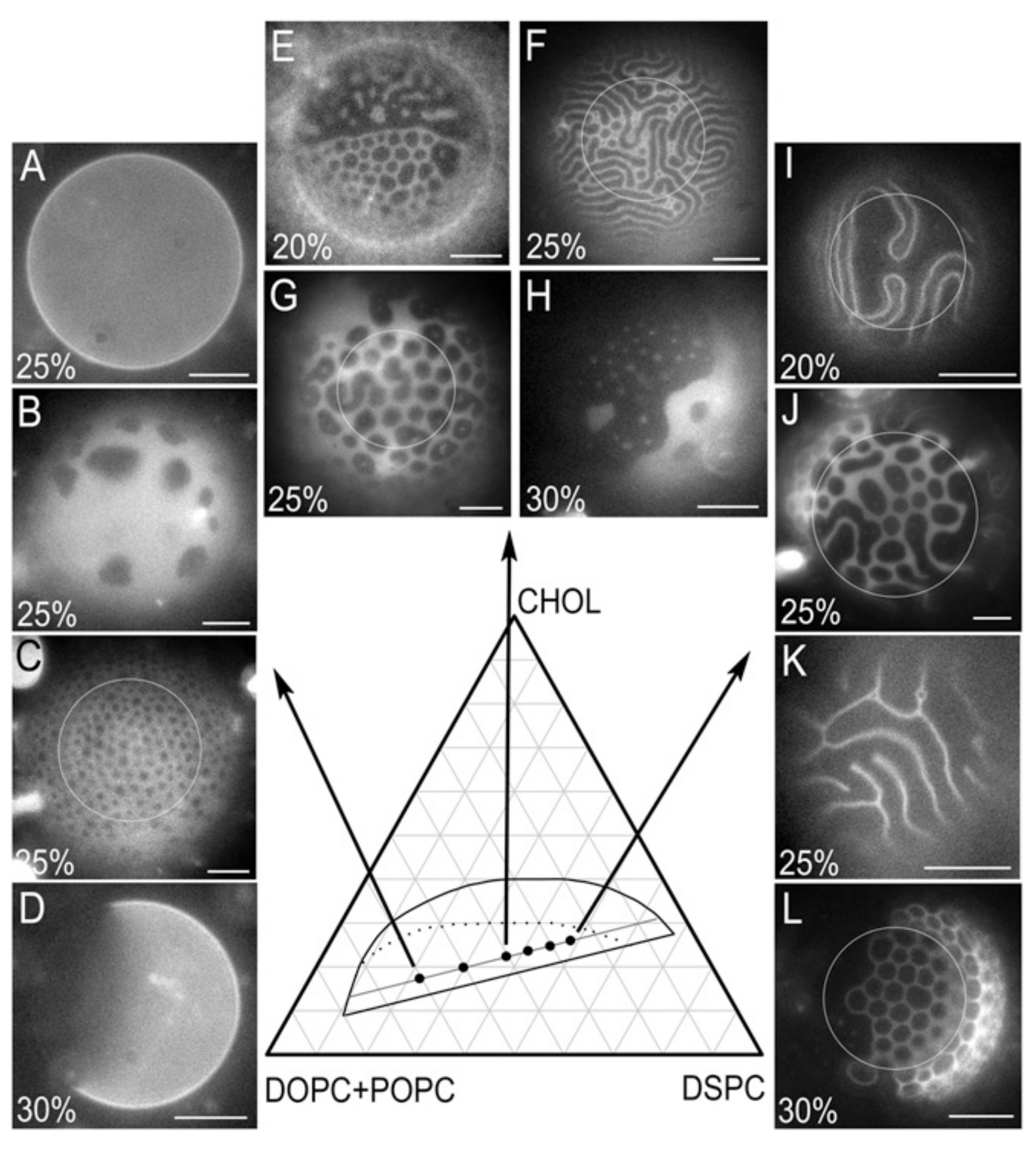} \\ \centerline{(c)}}
\end{tabular}
\end{center}
\caption{(\textbf{a})~Numerical model of biphasic vesicle with Lo (in white) and Ld (in red) separated phases, sampled by Monte Carlo simulations. The ratio $\kappa_{\rm Lo}/\kappa_{\rm Ld}$ is indicated below each vesicle. Most curvature is accumulated in the more flexible Ld phase. Reproduced from~\cite{Hu2011}, with permission from The Royal Society of Chemistry, Copyright 2011. 
(\textbf{b})~Another numerical modeling of a biphasic vesicle with various Lo and Ld area fractions (Lo is black and Ld is white), showing patterned morphologies. Panel (\textbf{a2}) is the same simulation snapshot as panel (\textbf{a1}), but displays the mean curvature map instead (curvature units in the color scale on the right are in $\mu$m$^{-1}$), thus illustrating the coupling between the local composition and curvature. The most flexible Ld phase is also the most bent one. Reproduced from reference~\cite{Amazon2013}, with permission from the American Physical Society, Copyright 2013. In both panels (\textbf{a},\textbf{b}), the spontaneous curvature $C_{\rm sp}$ has been set to 0 for both phases. (\textbf{c})~Fluorescence microscopy images showing a large variety of patterns in the 4-component 1,2-distearoyl-sn-glycero-3-phosphocholine(DSPC)/1,2-Dioleoyl-sn-glycero-3-phosphocholine (DOPC)/1-palmitoyl-2-oleoyl-sn-glycero-3-phosphocholine (POPC)/cholesterol
vesicles, displaying coexistence of Lo (dark grey) and Ld (light gray) phases below the demixing temperature $T_d$. The vesicles either appear homogeneous or display roundish domains or labyrinthine  phases according to the concentrations of the different lipids. The percentage indicated in each panel is the DOPC to DOPC+POPC molar ratio $\rho$. Scale bars: 10~$\mu$m. Temperature: 23 $^\circ$C. Reproduced from~\cite{Goh2013}, with permission from Elsevier, Copyright 2013.  }
\label{Feigenson:Fig}
\end{figure}

In two contemporary series of papers, Lipowsky and collaborators on the one hand, and Feigenson and collaborators on the other hand, have investigated in detail the role of the different bending elastic moduli between two separated phases, e.g., the Lo and Ld ones, and how they can lead to mesophase separation. These investigations combine experiments on GUVs, analytical considerations and numerical simulations (Figure~\ref{Feigenson:Fig}).

Lipowsky and his collaborators proposed two different approaches to clarify the role of two different bending moduli~\cite{Hu2011,Gutlederer2009}. Buckling as discussed in Section~\ref{buckling} is not at play here because the line tension $\lambda$ is below its critical value $\lambda_c$. The Lo and Ld phases can have a non-zero spontaneous curvature $C_{\rm sp}$, but this is not a prerequisite for the competing-interactions mechanism under consideration here. As above, the reduced vesicle volume $v$ can be either controlled by the osmotic pressure difference across the membrane, or free to fluctuate, thus~assuming in both cases the \mbox{membrane to be} permeable to water~\cite{PhillipsBook}. In 2009, Gutlederer et al.~\cite{Gutlederer2009} studied this problem by \mbox{appealing to a} numerical solution of the mechanical energy minimization problem, again ignoring thermal fluctuations and thus configurational entropy. A morphology diagram was obtained, \mbox{with or without} volume constraint. The configuration minimizing the interfacial energy at the phase boundary~is the macrophase separation, with a single domain of each phase. \mbox{However, when decreasing} the line tension $\lambda$ or increasing the $\kappa_{\rm Lo}$ to $\kappa_{\rm Ld}$ ratio or the Lo area fraction $\phi_{\rm Lo} \in [0,1]$, it was found that macro-domains split into two or more smaller ones in order to deal with the bending energy extra~cost. An additional volume constraint through the parameter $v$ leads to additional morphological transitions~of multi-domain vesicles, in particular oblate-to-prolate ones. To study the role of thermal fluctuations in this context, Hu, Weikl and Lipowsky later performed Monte Carlo \mbox{simulations on a} model of dynamically triangulated biphasic vesicles below the demixing temperature $T_d$~\cite{Hu2011} (Figure~\ref{Feigenson:Fig}a), where each triangle is either Lo or Ld. Adjacent triangles interact through an Ising-like model. By construction, Monte Carlo simulations fully take thermal fluctuations into~account.  In thermodynamic equilibrium, the obtained morphology diagram is even richer than the previous~one, with macro-domains splitting in up to 6 smaller ones, so that most curvature is absorbed by the more bendable Ld phase. Here also, constraining the reduced volume $v$ can have noticeable~effects. For~$v$ very close to 1, the vesicles are forced to adopt nearly spherical shapes with the same curvature for both~phases. They display a macrophase separation to minimize the interfacial energy. As~before, when~decreasing $v$, the vesicles undergo morphological transitions to reduce the bending energy in spite of the increased interfacial cost. Some morphologies predicted by this approach are successfully compared to experimental images~\cite{Veatch2003,Gudheti2007}. These morphologies have also been explored numerically by solving the two-fluid Stokes equation with a thermal noise in a two-dimensional curved surface~\cite{Taniguchi2011}. It is observed that several shapes can be in competition and that the final (meta-)stable shape may depend on the system history.

The first paper of the series co-authored by Feigenson on the same topic~\cite{Konyakhina2011}, in 2011, examined the experimental phase-behavior of 4-component DSPC/DOPC/POPC/cholesterol lipid mixtures~by fluorescence microscopy (POPC is 1-palmitoyl-2-oleoyl-sn-glycero-3-phosphocholine and DSPC is 1,2-distearoyl-sn-glycero-3-phosphocholine). A rich variety of patterning morphologies was observed on GUVs, comparable to what is shown in Figure~\ref{Feigenson:Fig}c, such as roundish small domains, honeycomb arrangements, stripes, or labyrinths. The main finding of this work is that mesophases are observed provided that a fourth lipid is added to the mixture. The DOPC to DOPC+POPC 
 molar ratio, denoted by $\rho$, \mbox{is to be in a} relatively narrow window to avoid either a macrophase separation or an homogeneous mixture (at~least homogeneous as it appears above the Rayleigh resolution limit). This patterning propensity~of this 4-component lipid mixture was explored further by the same group in Refs.~\cite{Heberle2013,Goh2013,Amazon2013,Amazon2014}, using~both Monte Carlo simulations and performing a more systematic exploration of the parameter~space. The reason why a fourth lipid species is required to observe mesophase separation in this context presumably comes from the lower line tension $\lambda$ between the Lo and Ld phases. Line tension has been shown to increase with $\rho$ because the thickness mismatch between Lo and Ld phases has been shown~to grows linearly with $\rho$ thanks to SANS experiments~\cite{Heberle2013}. Thickness mismatch goes from $0.6$~nm when $\rho=0$ to $1.1$~nm when $\rho=100$~\%. The line tension is then estimated to be a fraction~of pN~in the range~of values of $\rho$ where modulated phases are observed. When it grows beyond a limiting~value, in~the pN range, the interfacial energy is too large and the macrophase is more stable than the patterned mesophase in the GUVs of interest. In contrast, in the small 60~nm-vesicles of reference~\cite{Heberle2013}, 
a large domain presumably remains unstable even when $\rho$ grows because its bending energy remains large as compared to several smaller domains. A moderate line tension also forbids buckling~as~above. 

The simulations~\cite{Amazon2013} are based upon a model comparable to the one discussed just above, with~the vesicle discretized in small triangles. In this case however, the concentration field is defined~on the lattice vertices, contrary to Lipowsky's model where it is defined on the triangular faces. The~system~is also simulated below $T_d$ (Figure~\ref{Feigenson:Fig}b). The initial vesicle radius is taken to be $R=25$~$\mu$m and the volume~is not constrained, but non-concave shapes are forbidden. Even though the bending moduli $\kappa$ are $\sim10$ times higher than measured ones (up to $5000\,k_{\rm B}T$ for $\kappa_{\rm Lo}$), the model reproduces well the variety of morphologies observed in experiments. The line tension is also relatively weak, $\lambda \sim 10^{-2}$~pN, which indicates in principle the proximity of a critical point~\cite{Honerkamp2009}. Moderately increasing the line tension makes domains larger. Values of $\lambda$ in the pN range again lead to macro-phase separation instead of meso-patterning. This work also demonstrates that increasing both the vesicle radius $R$ and the line tension $\lambda$ while keeping $R/\lambda$ constant keeps morphology unchanged. In a subsequent experimental work~\cite{Goh2013} complementing Ref.~\cite{Konyakhina2011}, the parameter space is explored following tie-lines~in the coexistence region of the phase space (a tie-line is such that the composition of each phase, \mbox{Lo or Ld}, is fixed along it~\cite{veatch1}). The ranges of the DOPC to DOPC+POPC ratio $\rho$ for which a modulated phase can be observed have been characterized thoroughly along the tie-lines. The agreement between experimental and numerical morphologies is confirmed in this study, at least on a qualitative level. 

This series is continued by a fourth paper~\cite{Amazon2014} where a new ingredient is added in the simulations, namely dipole-dipole repulsion between lipids with a 1 to 3~nm range depending~on the solution ionic~strength. This repulsion was not included so far because the membrane was \mbox{coarse-grained at a larger}, micrometric length-scale. The dipole density is assumed to be about twice~as large in the Lo phase as in the Ld one. This model belongs to the competing-interactions class because the repulsion is antagonist to the lipid-lipid attraction concretized by the finite line tension below $T_d$~\cite{Seul1995}. Small domains of a size $> 10$~nm are observed in the simulations. As anticipated, the~higher the dipole density in the Lo phase, the smaller the domains. Vesicle modeling embracing both length-scales (nanoscopic and micrometric) remains to be done because this study was limited to patches of 60~nm due to the necessary low level of coarse-graining (see the Section~\ref{needed:theory} for further discussion on this~topic). This work also studied how the line tension is renormalized in the Ising universality class  when changing the coarse-graining level, but the coupling to membrane fluctuations~is still lacking \mbox{in this respect}. 

\begin{figure}[h]
\begin{center}
\ \includegraphics*[width=10cm]{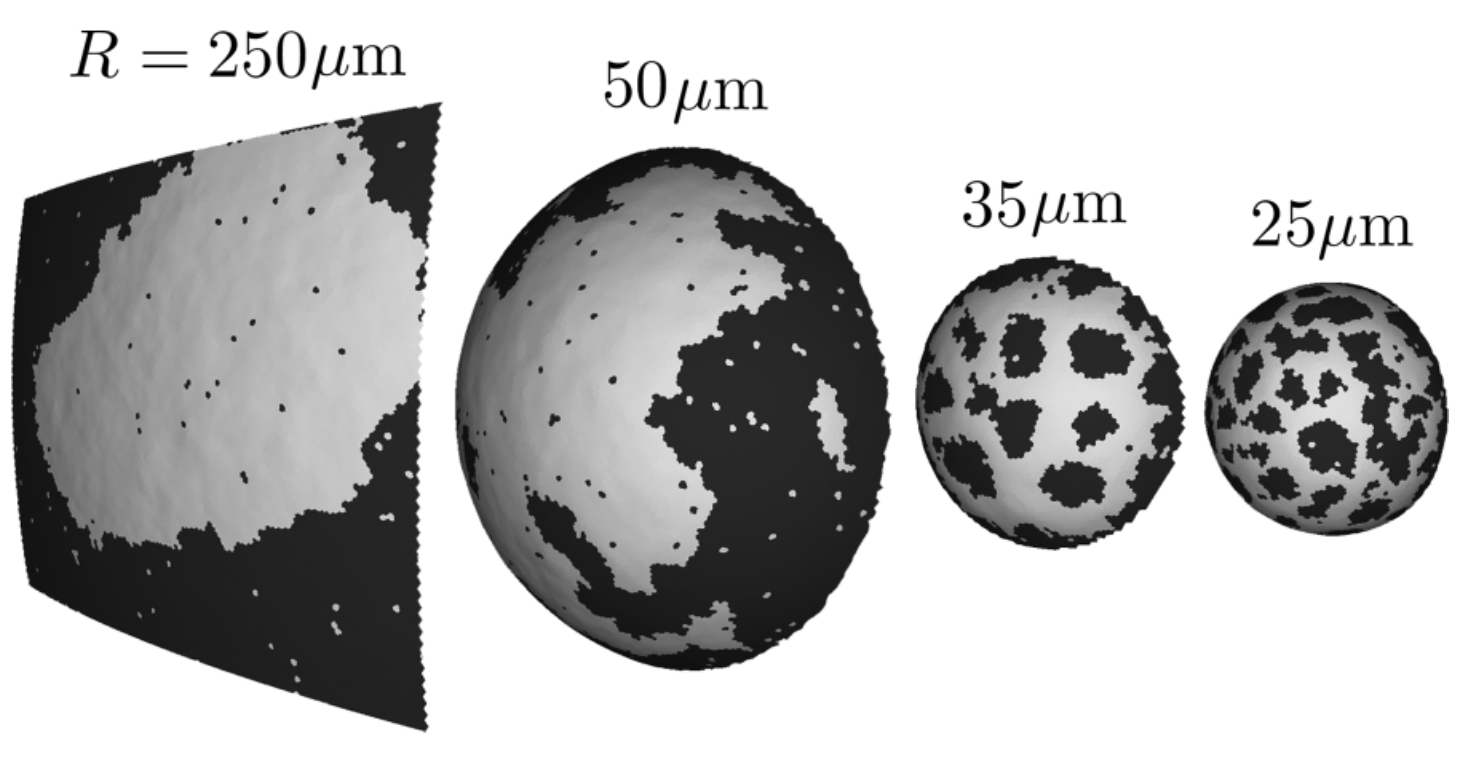}
\end{center}
\caption{From right to left, the vesicle radius of curvature (written above each snapshot) is increased  while the total area of the patch is held fixed. Above a critical curvature, there is a transition from a Lo/Ld mesophase to a macrophase separation. Reproduced from~\cite{Amazon2014}, with permission from the American Physical Society, Copyright 2014. }
\label{Carapace:Fig}
\end{figure}

Coming back to the situation without dipole-dipole repulsion, maybe the main finding of this work~\cite{Amazon2014} is that ``background curvature is necessary for the stability of patterns'', thus pointing out the explicit up-down symmetry breaking mechanism. Considering membrane patches with the same simulations parameters except an increasing radius of curvature $R$, it is observed that first domains broaden and eventually disappear to the profit of a macro-phase separation (when~$R \gtrsim 50$~$\mu$m), as~seen~by inspection of Figure~\ref{Carapace:Fig}. Indeed, as in Lipowsky and coworkers' studies cited above, bending the most rigid Lo phase requires elastic energy. Missing spontaneous curvature, this~rigid Lo ``carapace'' must be broken up into smaller fragments in order to conform to the curved spherical shape. The~interstice between the Lo domains is filled by the more bendable Ld phase, which minimizes the elastic energy. This theoretical conclusion seems to be confirmed by AFM experiments on a 4-component lipid mixture containing both POPC and DOPC, as above (DSPC has been replaced by brain-sphyngomyelin (bSM))~\cite{Ho2016}. This systematic study points out the absence of ``isolated disk-like domains'' in flat geometry (where $R \rightarrow \infty$) resulting from phase-dependent bending moduli, even at the 10~nm length-scale.

To confirm the decisive role played by line tension in the coexistence region below the demixing temperature on GUVs, a very recent work by Feigenson and coworkers~\cite{Usery2017} explored a large variety~of 4-component mixtures of the (high-$T_m$ PC or sphingomyelin)/low-$T_m$ PC(s)/cholesterol~type. In~particular, the bSM/DOPC/POPC/cholesterol mixture of~\cite{Ho2016}, 
as just mentioned above, is~also included in this study and does display mesophases on GUVs, thus confirming our previous statement. This work also refines further the parameter range, notably the value of $\rho$, where mesophase separation~is visible by optical microscopy (with respect to either homogeneous phases or macrophase separation). The line tension of the Lo/Ld interface was measured using the flickering spectroscopy method and confirmed to dwell in the 1~pN range. More precisely, patterned morphologies were observed for moderate values of the line tension $0.3 \lesssim \lambda \lesssim 1.1$~pN. The possibility that thermal fluctuations break up domains below 0.3 pN is evoked. Above $\approx 1.1$~pN, macrophase separation in large round domains is found to be favored.

{\subsection{Competing Interactions: Spontaneous Curvature Induced by Membrane Inclusions}
\label{cl:ph}}

{The approaches in Section~\ref{belowTm} were} focussed on the competition between line tension on the one~hand, and bending elastic energy, through the dependence of the bending rigidity $\kappa$ (and possibly the saddle-splay modulus $\kappa_G$) upon the local membrane local composition $\phi$ on the other hand. As~already discussed in Section~\ref{aboveTm}, another local parameter likely depends upon the local membrane phase state or local composition, namely the spontaneous curvature $C_{\rm sp}$. As seen in Section~\ref{curv:comp:cp}, several mechanisms can lead to local spontaneous curvature $C_{\rm sp} \neq 0$ of lipid bilayers. As far as membrane inclusions are concerned, their molecular conformation, for example, can also confer them~a conical shape that locally bends the membrane and thus plays a similar role. We shall mainly insist~on this second source of spontaneous curvature in the present section. {This section is not included in the previous strong segregation limit  one because it deals with membrane inclusions such as proteins instead of lipid phases. However, the effective theories proposed in this context often consider protein domains with little fluctuating boundaries, i.e., large line tension, and connections will be made with continuous theories as above.}

The basic mechanism is that the membrane can deal with the curvature constraint by splitting the macro-phase into small domains, such that regions with a strong spontaneous curvature $C_{\rm sp}$ are dispersed throughout the membrane. This will be shown to lower the overall bending energy.

\bigskip

\subsubsection{Inter-Protein Short-Range Forces}

Before dealing with the role of spontaneous curvature in the stabilization of mesophases, we~need~to briefly review the origins, ranges and intensities of inter-protein attractive forces that lead~to their phase-separation in biomembranes, and the interplay with lipidic phase-separation. \mbox{Indeed, the fact} that they are inserted in (or adsorbed on) the membrane is the source of \emph{effective} 
 interactions (also called ``potential of mean force'') propagated by the elastic and fluctuating membrane. They can be either of mechanical origin, or of entropic nature, resulting from the important thermal fluctuations of the membrane. Many works have calculated these forces propagated by the membrane, for example the pioneering works of references~\cite{goulian,fournier,evans,Park1996,Gil1997} and many of those cited below, using the Helfrich free energy to describe the membrane~\cite{helfrich1973} and different kinds of boundary conditions imposed by the inclusions.

\medskip

\begin{figure}[h]
\begin{center}
\ \includegraphics*[width=14cm]{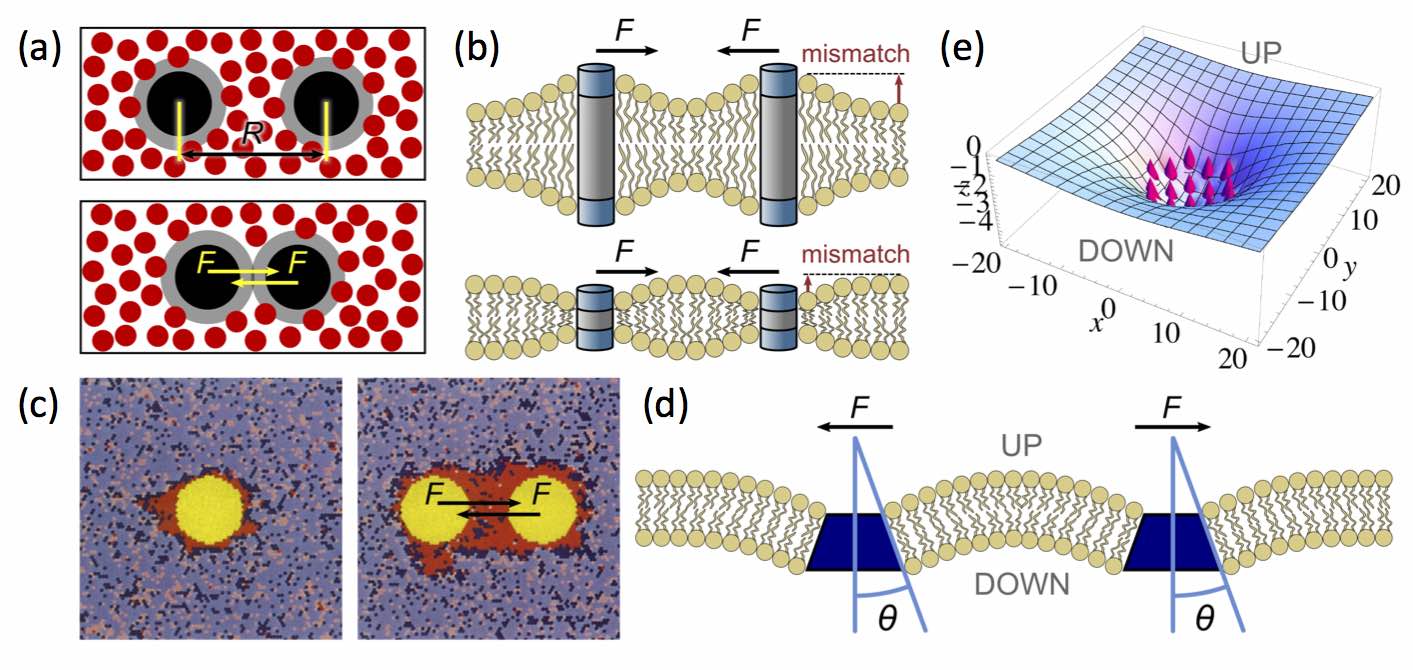} \ 
\end{center}
\caption{Illustrations of some of the inter-protein forces (denoted by ${F}$ on this figure) propagated by the lipid membrane. (\textbf{a})~Depletion forces due to the lateral osmotic pressure exerted by lipids (red discs) on the proteins (black). The pressure becomes anisotropic when the distance between proteins is smaller than the lipid size.   (\textbf{b})~Hydrophobic mismatch forces resulting from the difference between the thickness of the membrane hydrophobic layer (the hydrocarbon chains) and the height of the protein hydrophobic cores (in gray) (\textbf{c})~Wetting-induced interaction: when the membrane is made of a lipid mixture, a single protein (left panel) nucleates a ``halo'' of ``wetting'' lipids (in red). When two proteins get closer (right panel), their halos overlap, which reduces the interfacial energy between lipid species. A force ensues. (\textbf{d})~Mutual interaction felt by two up-down non-symmetric inclusions (in blue) and propagated by the deformable elastic membrane. The asymmetry is schematized by the conical shape of inclusions. It is measured by the cone half-aperture angle $\theta$. (\textbf{e})~Collective deformation of the membrane by an assembly of up-down non-symmetric inclusions (pink cones). Figures reproduced from~\cite{Destain2016}, with permission from Elsevier, Copyright 2016.}
\label{Lang:Fig} 
\end{figure}

\begin{enumerate}
\item \emph{Electrostatic, van der Waals and hydrogen-bond interactions}---Polar and charged amino acids at their surface can interact when two proteins come in close proximity. The Debye length $\lambda_D \sim 1$~nm in water at physiological salt condition~\cite{PhillipsBook} sets the typical range above which these interactions~are~screened. Inside the apolar hydrophobic membrane region where the dielectric constant~is~weaker, the range can be somewhat larger, of a few nanometers~\cite{Abney1993,liu2005}. The range~of van der Waals and hydrogen-bond interactions is also nanometric. 
\item \emph{Hydrophobic mismatch}---Integral proteins have transmembrane domains consisting of alpha helices with hydrophobic amino-acid side chains, buried inside the hydrophobic core of the lipid membrane. Protein and membrane hydrophobic core thicknesses do not necessarily match. Since exposure of hydrophobic residues to the aqueous solvent is energetically unfavorable,  the membrane must be deformed in the case of significant mismatch~\cite{Mouritsen2005,Killian1998}. If two (or more) proteins are in proximity, the overall energy penalty depends on their distance $d$. As above, an~effective force ensues (Figure~\ref{Lang:Fig}b). It is attractive when both mismatches have the same sign and repulsive in the converse case. The energies at play go from a fraction of $k_{\rm B}T$ to several~$k_{\rm B}T$, depending on the degree of hydrophobic mismatch, and the range of these forces~is few nanometers~\cite{Bohinc2003,Schmidt2008,Yoo2013,Bories2018}. It has been suggested that hydrophobic mismatch forces are not pairwise additive~\cite{Brannigan2007}.
\item \emph{Casimir interaction}---This attractive interaction, of entropic origin, is named by extension of the Casimir interaction in quantum physics (the attraction between conducting plates mediated~by quantum fluctuations in the electromagnetic field). Here it results from the transverse thermal fluctuations of the elastic membrane. The number of vibrational degrees of liberty of a membrane~in which two (or more) inclusions are embedded depends on their mutual distance $d$. The potential of mean force thus depends on $d$, and has been shown to behave as $ \sim -  k_{\rm B}T / d^4$~in the case of vanishing membrane tension $\sigma$~\cite{Park1996,Dommersnes1999}. The calculation can be extended to the case $\sigma>0$, where the interaction energy decays much faster with $d$, as $\sim -  k_{\rm B}T / d^8$ when $d \gg \xi$, and~as $ \sim - k_{\rm B}T\exp(-d/\xi)$ when $d \lesssim \xi$~\cite{Lin2011,Weitz2013}.
\item \emph{Depletion (or excluded-volume) forces}---Attractive depletion forces (Figure~\ref{Lang:Fig}a) are well characterized~in soft condensed matter when large particles evolve among smaller ones, and~play~a role in physical biology (see~\cite{PhillipsBook} for example). In the present case, they~are due~to the 2D osmotic pressure laterally exerted by the surrounding lipids on large transmembrane proteins (larger than lipids). It should be far less pronounced for peptides. Roughly~speaking, when two proteins are far away, the lateral osmotic pressure is isotropic and no net force~ensues. When the relative distance becomes on the order of the lipid lateral size ($<1$~nm), the~interval between the two inclusions tends to be depleted in lipids, and the pressure is not isotropic~anymore. This tends to bring proteins closer when they are about a nanometer away~\cite{Sintes1997,Borodich2003}. The ensuing binding energy is on the $k_{\rm B}T$ range, even though the actual value depends on the model details. 
\item \emph{Lipid wetting}---Some lipids are known to have a preferential affinity for given proteins species~\mbox{\cite{Hancock2006,Hinderliter2004,Reynwar2009,vandenBogaar2011}}, in particular but not exclusively because they better match their hydrophobic length. Even above the phase-transition temperature, the protein can nucleates a small ``halo'' of such lipids, the range of which is on the order of magnitude of the composition correlation length $\xi_{\rm OZ}$ (see Section \ref{equil}). This mechanism known as ``wetting''~\cite{Gil1997,Gil1998a,Gil1998b}~is reminiscent of the ``lipid annulus'' or ``lipid shell'' concepts that have become popular~in the biophysical literature a dozen of years ago~\cite{Poveda2008}. When two proteins approach close enough for their halos to overlap, they tend to assemble because it reduces the net interfacial~energy. An~effective attractive force ensues (Figure~\ref{Lang:Fig}c). This nucleation mechanism can also promote the formation of a lipid halo of a thermodynamic phase that would be unstable in absence~of~the~inclusion.  A similar mechanism has been demonstrated to emerge in a very illustrative way~\cite{Katira2016}. In all cases, the range is set by the correlation length $\xi_{\rm OZ}$. 

\quad\quad This force is enhanced near a miscibility critical point because the composition correlation length $\xi_{\rm OZ}$ grows significantly. Exactly at the critical point,  a long-range, power-law decrease of the potential of mean force at large inter-inclusion distance $d$ has been predicted by a conformal field theory approach, with exponent $-1/4$, and confirmed by Monte Carlo simulations of the Ising model~\cite{Machta2012}. Coarse-grained molecular dynamics simulations on a model membrane and~a phenomenological Ginzburg-Landau theory have explored the same mechanism in the case~of peripheral proteins adsorbed onto the bilayer and interacting preferentially with one lipid species (among two). They drawn similar conclusions~\cite{Reynwar2009}. The binding energy at close range for two identical particles is also found in the $k_{\rm B}T$ range.

\quad\quad Note that this mechanism is specific to the protein species and the lipids with which~it preferentially interacts because the halos must be miscible if the interaction is attractive. \mbox{In the case} where they are immiscible, the force can even become repulsive instead~\cite{Reynwar2009,Machta2012}. Small alterations in lipid chemical structure can thus lead to dramatic changes in the membrane~organization. This mechanism has been evidenced in model membranes~\cite{Hinderliter2004}. 
\end{enumerate}

The majority of these forces mediated by the lipid membrane are attractive, notably the ubiquitous Casimir and depletion ones, which explains why proteins are found in small clusters in bio-membranes. However, the variability of protein structures and the combination with possibly repulsive forces likely tune the strength of the attraction in function of the interacting proteins. It will be used in point Section~\ref{pt5} below. 
 Since the intensity of each force is on the order of $k_{\rm B}T$, the typical binding energies $\epsilon_{\rm att}$ between membrane proteins are a few times larger than the thermal energy $k_{\rm B}T$, and are sufficient~to drive protein macro-phase separation if no additional forces counter-balance them at long~range. This~order~of magnitude has been confirmed by refined numerical simulations~\cite{Schmidt2008,Reynwar2009,deMeyer2008,Periole2012}. High-precision AFM experiments have been able to confirm the orders of magnitudes of range and intensity of these forces in the case of ATP-synthase c-rings~\cite{Casuso2010}. The range was found to be $\approx 3$~nm and the binding energy $\epsilon_{\rm att} \approx 3.5 k_{\rm B}T$.

\subsubsection{The Cluster Phase Scenario}
As introduced at the beginning of this Section~\ref{equil}, cluster phases are patterned structures resulting from a competition between strong short-range attraction and longer-range, weaker repulsion between proteins~\cite{Stradner2004}. Thus this mechanism also belongs to the competing-interactions model class~\cite{Seul1995}. It is reminiscent of the micellization theory~\cite{SafranBook}: domains, named clusters here, have a non-extensive free energy because of the long-range repulsion. 

To make the argument more precise, we write a functional form of the free energy $F(k)$ of a cluster containing $k$ proteins, supposed to be identical in a first step~\cite{DestainForet2008}:
\begin{equation}
\label{Fk}
F(k) = -f_0(k-1) +  \rho_0 (k-1)^{\frac{d-1}{d}} + \chi_0 (k-1)^{\alpha},
\end{equation}
with $f_0,\rho_0,\chi_0>0$ and $\alpha>1$. The first term of $F(k)$ is its extensive part, where $f_0$ is the energy \mbox{gained by~a} protein when it joins the cluster. It is proportional to the short-range protein-protein binding energy $\epsilon_{\rm att}$, as discussed above. The second term accounts for the \mbox{proteins at the} cluster~boundary, that are energetically disfavored as compared to those in the cluster~bulk. The~coefficient $\rho_0$ is thus proportional to the cluster line tension $\lambda_{\rm cl}$, \mbox{itself related to $\epsilon_{\rm att}$}. Finally,~the~last term takes long-range repulsion into account in an effective way, by growing faster than the cluster~size~$k$. \mbox{In~the extreme} case where each protein pair inside the cluster experiences the same repulsive energy $\epsilon_{\rm rep}$, the total repulsive energy is $\epsilon_{\rm rep}\, k^2/2$, thus $\alpha=2$~\cite{Wasnik2015}; $\chi_0$ measures the \mbox{intensity of the repulsion}. This expression is the so-called ``droplet approximation'' with an additional non-extensive term (i.e., a term growing faster than $k$). Note that in Equation~\eqref{Fk}, $F(k)$ has been written~as~a function~of $(k-1)$ so that $F(1)=0$, as desired. This is a technical choice that has no incidence~on the statistical mechanics of clusters for which $k^* \gg 1$. Replacing $(k-1)$ by $k$ and subtracting a constant~to ensure that $F(1)=0$ leads to the same conclusions.  
This term makes too large clusters unstable because they have~an unfavorable energy cost, and leads to the observed mesophase separation~\cite{SafranBook}, as demonstrated {via} Monte Carlo simulations~\cite{Destain2008}, analytical calculations based~on the principles of statistical mechanics~{in the limit where interactions between clusters are negligible \mbox{(i.e., in the dilute limit)}}~\cite{DestainForet2008}, or more phenomenological arguments~\cite{Sieber2007,Gurry2009}.

As a result, above a protein threshold concentration $\phi_c$, the cluster size distribution is bimodal. A~protein dilute ``gas phase'',  constituted of protein monomers and rare dimers or trimers, coexists~with large multimers, or clusters, with a typical protein number $k^*$ depending on the balance between attraction and repulsion. Just above $\phi_c$,  one finds that $k^* \propto (\rho_0/\chi_0)^{(\alpha-1/2)^{-1}}$~\cite{DestainForet2008}. This~scaling can~be simply obtained by balancing the second and third terms of $F(k)$ in Equation~\eqref{Fk}, accounting~respectively for short-range attraction, via the line tension, and longer-range repulsion.  The typical size $k^*$ of multimers, excluding 
 monomers, is expected to grow only slowly with $\phi$ above $\phi_c$~\cite{DestainForet2008}. This is consistent with experimental observations on membrane protein clusters, both in vitro~\cite{Gulik1987} and in vivo~\cite{Whited2015,Homsi2014,Merklinger2017}. In contrast, in a wide range of parameters, it is also found that the average cluster size $\langle k \rangle$, including 
  monomers, is proportional to the protein concentration: $\langle k \rangle \propto \phi$ above $\phi_c$~\cite{Destain2008,DestainForet2008}. This is also consistent with soft matter experiments~\cite{Stradner2004}. 

Note that in this scenario, the clusters are seen as dynamical entities, with a well-defined~boundary, but that however, permanently exchange isolated molecules with the surrounding membrane \mbox{(the above ``gas phase'')}~\cite{Honigmann2014,Destain2008}, because the binding energies at play are on the same order of magnitude as the thermal energy $k_{\rm B}T$. Experimental observations have confirmed this prediction~\cite{Espenel2008}, showing with the help of dual color fluorescence microscopy and particle tracking how a single protein enters a cluster, dwells in it for a duration of several seconds, and eventually leaves~it~to diffuse again freely in the cell membrane. This allows a rapid and efficient remodeling of clusters, in~function of the evolution of inter-protein interaction parameters, themselves depending on protein conformational~changes, e.g., upon receptor activation. For example, in Ref~\cite{Rosholm2017}, it has been measured in living cells that the spontaneous curvature of some G-Protein Coupled Receptors (GPCR) significantly changes upon ligand binding. The close lipidic environment can also depend~on the protein conformation~\cite{Harder2003}. Dynamical re-targeting to different clusters after activation can \mbox{then be envisaged}.

\subsubsection{Spontaneous Curvature Can Play the Same Role as a Long-Range Repulsion}

When two up-down asymmetric inclusions locally bending the membrane are present at a given distance $d$, they feel a mutual interaction propagated by the elastic membrane (Figure~\ref{Lang:Fig}d). As~compared~to the Casimir attraction evoked in Section~\ref{bending}, it comes from the mechanical deformation of the membrane and not from thermal fluctuations~\cite{goulian,Weitz2013,Fournier1997,Weikl1998}. The interaction is repulsive at long range if both inclusions have the same orientation, and attractive in the converse case where they are ``head-to-tail''. The potential of mean force falls off as $d^{-4}$ at large $d$ in the tensionless~case, and exponentially as $\exp(-d/\xi)$ for a taut membrane (in fact Bessel functions~as~in Equation~\eqref{VdeR}, but with a different prefactor), with a range set by the Helfrich correlation length $\xi=\sqrt{\kappa/\sigma}$. The intensity of the interaction is proportional to $\theta^2 \sigma$, where $\theta$ is the cone half-aperture angle as illustrated in Figure~\ref{Lang:Fig}d. Note that for a protein of radius $a$ in the membrane plane, $\theta$ and $C_{\rm sp}$ are simply related through $\theta = a \, C_{\rm sp}$.

When an assembly of such identical proteins dwell in close proximity, they collectively bend the membrane and provoke an invagination as shown in Figure~\ref{Lang:Fig}e. The situation becomes more involved because the energy cannot be written as a sum of pairwise additive interactions anymore. This question has been studied in details in Ref.~\cite{Weitz2013}, by calculating exactly the elastic deformation shape and its energy cost. The main finding is that the elastic energy of a $k$-cluster can be written in the form of Equation~\eqref{Fk}, with $\alpha \approx 1.5$. It follows that the spontaneous curvature imposed by \mbox{proteins in a} cluster is formally equivalent to a long-range repulsive force when expressing the free-energy $F(k)$ as in Equation~\eqref{Fk}. As a long-range repulsion would do it, it makes large cluster unstable because the membrane deformation elastic energy grows faster than $k$ and overcomes the gain in terms of line tension if $k$ is too large. 

A simple argument has been proposed in~\cite{Weitz2013} to account for this non-extensivity. In~the low-surface tension regime $\sigma \ll \kappa/R^2$, where $R$ is the vesicle radius, Helfrich's free energy is dominated by the curvature term. The tension term can be treated as a perturbation. The spontaneous curvature $C_{\rm sp}$ is imposed in the domain by the asymmetric inclusions. Minimizing Helfrich's free energy results, at order 0 in the perturbation, in a parabolic membrane invagination inside the cluster $\langle h(\mathbf{r}) \rangle=C_{\rm sp} r^2/2$ if the cluster is centered at the origin. Note that in the present approximation of small gradients~of~$h$, a~paraboloid of revolution is identical to a spherical cap. If $r_c$ is the cluster radius, Helfrich's free energy inside the cluster can be written
\begin{equation}
F(r_c) = \int_0^{r_c} \left(\frac{\sigma}2  C_{\rm sp}^2 r^2 + \frac{\kappa}2  C_{\rm sp}^2 \right) 2\pi r {\rm d}r 
 = \pi C_{\rm sp}^2   \sigma \frac{r_c^4}4 + \pi C_{\rm sp}^2 \kappa \frac{r_c^2}2 .
 \label{Fquad} 
\end{equation}

The first term is a correction to the second one because $\sigma$ is small. This term is proportional to $r_c^4$ and thus makes the total membrane free energy non-extensive, as required. Since the cluster number~of proteins satisfies $k\propto r_c^2$, we infer that $\alpha=2$ and $\chi_0 \propto C_{\rm sp}^2   \sigma \propto  \theta^2 \sigma$ in Equation~\eqref{Fk}. When~$\sigma$ takes intermediate values more compatible with real cell tensions, the exponent $\alpha$ appears to be slightly lesser than 2. This effective repulsion term is always positive, even for weak asymmetry and weak~tension. 

A connection can be made with Section~\ref{belowTm}, where buckling of lipid domains was considered~{by following a continuous theory approach}. The study was extended to the case $C_{\rm sp} \neq 0$ by several authors~\cite{lipowsky,Ursell2009}. They naturally concluded that spontaneous curvature breaks the up-down symmetry and that budding then occurs in a preferred direction, regardless of the line tension value. This~is the same mechanism as above, even though tackled with a different mathematical approach. They~quantified how the spherical cap is deformed in this case when $\sigma$ grows. {A similar approach was proposed in parallel in Refs.~\cite{Harden1994,Harden2005} in the simplified case where all domains have the same size and form a regular two-dimensional array. It was concluded that the modulated phase is the most stable provided that the membrane tension is smaller than the same limiting value $\sigma_{\rm cr}$ as given~in Equation~\eqref{sigma:lim} (when both phases have the same curvature rigidity $\kappa$). It was also found that the number of domains grows proportionally to the concentration $\phi$~\cite{Harden1994}, and consequently that the typical domain size does not depend significantly on $\phi$, as already predicted by the cluster phase approach above.} In principle, the membrane elastic energy can be calculated in function {of the domain size and the statistical mechanics of nanodomains can be tackled, as in the above cluster phase scenario}. This remains to be done.

To our knowledge, the best available experimental test of this mechanism in a controlled way was performed by Shimobayashi and coworkers in 2016, by inserting GM1 gangliosides in the outer leaflet of GUVs~\cite{Shimobayashi2016}. GM1 gangliosides are known to insert preferentially in the Lo phase and then to impose a strong local spontaneous curvature to the membrane~\cite{Dasgupta2018}. Thus Lo domains are endowed with a local spontaneous curvature in this system, whereas Ld domains are not. First~the ternary DOPC/DPPC/cholesterol GUVs were prepared without GM1. The temperature was set below $T_d$. To~avoid buckling and the ensuing trapping in a metastable state (see Section~\ref{buckling}), the~experimental protocol ensured that the vesicles were close to spherical. Consequently, the Lo and Ld phases were fully separated. Then GM1 was added in the solution and got inserted in the outer~leaflet. \mbox{After several hours}, the macrophase separation disappeared to the benefit of a mesophase~in almost all~vesicles, thus proving experimentally that large domains with a spontaneous curvature are unstable. An interesting intermediary stripe morphology was observed before turning~to roundish small curved~domains, which were stable for several days. Higher concentration of GM1, thus higher spontaneous curvature lead to smaller domains, as expected.

\quad In a cell plasma membrane, there is no reason why proteins should preferentially curve the membrane in one direction, inward or outward. If one assumes that there are roughly as many inclusions imposing an inward spontaneous curvature as there are \mbox{bending it outward,} they~conceivably cancel each other out and lead to a globally vanishing membrane spontaneous~curvature. This simplistic argument has been refuted in reference~\cite{Weitz2013}. It has been shown~by using Helfrich free energy that mixing differently oriented inclusions in a same nanodomain is energetically unfavorable because it imposes geometric frustration on the height function $h(\mathbf{r})$. Consequently, inward and outward bending inclusions phase-separate and the cluster phase mechanism then applies~to each orientation, leading to distinct inward bent clusters and outward bent~ones. Such two differently oriented clusters feel an attraction at long range but a strong repulsion~at short range (when their separation becomes shorter than $\xi$)~\cite{Weitz2013}.

\subsubsection{Sources of (Local) Spontaneous Curvature $C_{\rm sp}\neq 0$}

We have listed at the beginning  of Section~\ref{curv:comp:cp} the variety of up-down-symmetry breaking mechanisms involving lipids. Proteins also come into play to locally curve the membrane in collaboration with lipids~\cite{Zimmerberg2006}, as follows: 
\begin{enumerate}
\item The transmembrane part of an integral protein has no reason to be up-down symmetric, not least because the cytosolic and extracellular protein regions do not have the same biological~function. This is either apparent in the molecular shape of transmembrane proteins or can be inferred from their behavior in biophysical experiments~\cite{Zimmerberg2006,Rosholm2017,Doyle1998,Bass2002,Park2008,Manglik2012,Aimon2014,Prevost2015}. However, it seems difficult to infer the spontaneous curvature from the sole molecular shape displaying up-down symmetry~breaking, for the reasons that we discuss now.
\item Peripheral proteins naturally break the up-down symmetry~\cite{Jarsch2016,Blood2006,Pezeshkian2016}, to a degree that depends~in particular on the depth of penetration of the hydrophobic domain of the protein into the bilayer~\cite{Zimmerberg2006}. Numerical evidence can be found for example in Ref.~\cite{Reynwar2009}, where the small shoulder on the interaction potential at intermediate range indicates a weak~repulsion. \mbox{More generally}, anchored molecules can play the same role, as it was non-ambigously demonstrated in reference~\cite{Shimobayashi2016} on experimental proofs.
\item The coupling between lipid composition and protein wetting by lipids is also a potential source~of local curvature if a protein recruits different lipids in the two leaflets, themselves promoting markedly differential local curvature of the two leaflets.
\end{enumerate}

Of course, local spontaneous curvature likely results from the interplay of these different mechanisms, notably in a nanodomain where several lipid and protein species cohabit.

\subsubsection{Diversity of Membrane Proteins and Biological Specialization of Clusters}
\label{pt5}
The above theoretical developments deal with a single protein species. In really, a cell plasma membrane usually contains hundreds of different populations that ensure as many biological functions. Clusters are thought to be specialized on the biological level, by gathering one or a few protein species out of the hundreds ones present in the membrane in order to perform a determined biological~function. In spite of their structural similarity, several membrane proteins have even been observed to segregate~in separate clusters~\cite{Homsi2014,Low2006,Henis2009,Milovanovic2015,Zuidscherwoude2015,Sajman2017,Houmadi2018}, thus enforcing the idea of an efficient, fine-tuned sorting~of membrane proteins in function of their mutual affinities. From~a statistical mechanics perspective, the~more natural way to account for these observations is to suppose that fine-tuning of protein-protein short-range attractions enables their targeting to different, specialized~clusters. In this respect, a~mechanism for specialization of clusters as been proposed as follows~\cite{Destainville2010,Meilhac2011}. We have seen above that even though some short-range forces are generic to (almost) all membrane proteins, such~as Casimir or depletion forces, most of them depend on the protein specificities (molecular shape, width of the hydrophobic core, surface charge density, physico-chemical affinity for specific lipids). Consequently, a very wide range of protein-specific modulation of the short-range affinity $\epsilon_{\rm att}$ can be confidently envisaged~\cite{Milovanovic2015,Plowman2008,Kaiser2011}.  

The concept of protein ``family'' can be introduced, a family regrouping the protein species that tend to co-localize in the same specialized clusters (e.g., a receptor, a G-protein and an effector)~\mbox{\cite{Destain2016,Remorino2017}}. We denote by $q$ the number of such families, by $\epsilon_{{\rm att},=}$ the binding energy (or affinity) between proteins~of the same family and by $\epsilon_{{\rm att},\neq}$ the binding energy between proteins of two different family. We introduce the difference of affinities $\Delta \epsilon = | \epsilon_{{\rm att},\neq} - \epsilon_{{\rm att},=}|$ that indicates the degree of preference that proteins have for partners of their own family. Appealing to a mean-field Flory-Huggins theory~\cite{deGennes1979}, one defines a new modified droplet approximation~\cite{Destainville2010}. For example, if $q=2$,
\begin{equation*}
\label{Fk2}
\begin{array}{c}
F(k_1,k_2) = -f_0(k-1) +  \rho_0 (k-1)^{\frac{d-1}{d}} + \chi_0 (k-1)^{\alpha} 
\end{array}
\end{equation*}
\begin{equation} 
 \begin{aligned}
+ \left[ \nu \Delta \epsilon \, x(1-x) + x\ln x + (1-x)\ln (1-x)  \right] (k-1)
\end{aligned}
\end{equation}
where $k_1$ and $k_2$ are respectively the numbers of family-1 and family-2 proteins in the cluster, \mbox{$k=k_1+k_2$, $x=k_1/k$ and $\nu$} is the average number of neighbors of protein in the cluster bulk ($\nu \simeq 6$~in~2D). The last terms account for the mixing entropy. The Flory-Huggins theory states that there exists a critical value $\Delta \epsilon_c$ below which the two families are mixed in the clusters. In contrast, if~$\Delta \epsilon > \Delta \epsilon_c$, a~majority color emerges in each cluster, as desired. This argument can be extended to any value of $q$ and it has been demonstrated that the critical value becomes exact in the large-$q$ limit in spite of the mean-field approximation:
\begin{equation}
\label{Flory}
\Delta \epsilon_c = \frac{2}{\nu} \, k_{\rm B}T \, \ln q .
\end{equation}

This value has been confirmed by Monte Carlo simulations up to $q=5$~\cite{Meilhac2011}. The main interest of this result is that the logarithm grows very slowly with $q$. Even if $q=10^3$ or even $10^4$ different protein families cohabit in the same cell plasma membrane, $\Delta \epsilon_c$ does not exceed $3~k_{\rm B}T$. We have seen in the previous paragraphs that such a modulation of the short-range affinity is easily attainable thanks to the different forces propagated by the membrane. 

We have thus shown that a ``smart'' mechanism for nano-sorting and cluster specialization is at hand without appealing to complex active processes. Of course, future studies will have to deal with the variability of the parameters  $\epsilon_{{\rm att},=}$ and $\epsilon_{{\rm att},\neq}$ with the proteins species at play, whereas we have considered so far that each of them can take only one value. 

{\subsection{A Unifying Rationale: Up-Down Symmetry Breaking}}
\label{cl:strong}

{In Sections~\ref{belowTm} and~\ref{cl:ph},} we have examined into detail the three main mechanisms leading to the splitting up of the macroscopic Lo phase below $T_d$ in order to balance the elastic energy and the interface length. Note that the configurational entropy is also increased when splitting the macrophase, which should in principle be included in more refined theories. 

Without any source of explicit up-down symmetry breaking, a strong line tension $\lambda$ well below the phase-separation temperature is sufficient to promote spontaneous symmetry breaking leading to two equally probable buckled configurations of the Lo domains, upward or downward, the precise shape of which depends on the applied surface tension $\sigma$. In the planar topology, the equilibrium configuration remains the complete macrophase separation, but metastable configurations with long lifetimes can~emerge. They are constituted of budded domains that feel mutual repulsion mediated~by the elastic membrane. In the spherical topology of GUVs, configurations with several domains instead~of a single macrophase have been observed, both experimentally and numerically.

We have also examined two mechanisms where the up-down symmetry is explicitly broken, theoretically predicted and experimentally observed. On the one hand, even in the case of an up-down symmetric membrane, the symmetry can be explicitly broken because of the finite curvature imposed in the spherical topology. Like a ``carapace'' forced to be curved, the Lo phase is broken into smaller pieces when it is bent because it is more rigid than the surrounding Ld phase. An equilibrium mesophase separation emerges.

On the other hand, membrane inclusions (such as proteins) or asymmetric lipid composition~of both leaflets can lead to local spontaneous membrane curvature $C_{\rm sp}(\mathbf{r}) \neq 0$ coupled to the local composition $\phi(\mathbf{r})$. In this case, the elastic energy of the domain grows \mbox{faster than its area} \mbox{(it is non-extensive),} which sets a maximum domain size above which domains become thermodynamically unstable. In this case also, a patterned mesophase is stable in equilibrium, also~called a cluster phase in the case of membrane proteins.

These three mechanisms can be compared to the ones presented in Section~\ref{curv:comp:cp} in the weak segregation limit, even though quantitative comparison is difficult because the technical details are not always easy to put in direct correspondence between the different models. No spontaneous symmetry-breaking mechanism was identified in the weak segregation limit where the line tension is too weak to induce buckling. In contrast, differences in rigidities $\kappa$ or in spontaneous curvatures $C_{\rm sp}$ have been shown to lead to rich phase diagrams in both limits, especially in the spherical geometry. As~compared to the strong segregation case, the weak segregation one proposes, in addition to mesophases, the existence of SD phases where density fluctuations are transient with a typical size corresponding to a maximum of the structure factor. Density fluctuations are known~to exist in any biphasic system, with a typical size set by the Ornstein-Zernike correlation length $\xi_{\rm OZ}$ in absence of coupling. Because shape and composition are coupled, SD-phase fluctuations are different since their size is proportional to the Helfrich correlation length $\xi$, characteristic of shape~fluctuations. It is the natural length since at distances larger than $\xi$, the membrane mechanics~is only controlled by the surface tension $\sigma$ and bending effects become negligible. This is illustrated by Equations~\eqref{qc1}~or \eqref{qc2} for coupling to spontaneous curvature and detailed in~\cite{Gueguen2014} for coupling to bending rigidity. Presumably, their typical size ensues from the same physical mechanism as in the mesophases, as follows. In the vicinity of a miscibility critical point, critical phenomena theory predicts that the domain size distribution decays with a power law cut off at a size set by $\xi_{\rm OZ}$~\cite{Binder1986}. But too large domains, even though transient, should become energetically unfavorable in the coupled case for the same reason as in the strong segregation limit, because of their increased elastic energy induced by either $\kappa_1$ or $C_1$ (or both). They split into smaller domains, the typical size of which is now set by $q_c^{-1}$.

In the vicinity of the miscibility critical point,we have seen that $\kappa_1$ does not play any role in the planar case at Gaussian order, whereas it does in the spherical case. This is consistent with the findings~in the strong segregation limit, where the role played by $\kappa_1$ is directly related to the fact that $R$~is finite (see Figure~\ref{Carapace:Fig}). More quantitatively, if only the dependence of $\kappa$ with $\phi$ is considered~in this spherical geometry, and not the spontaneous curvature, then one must focus on the $C_1=0$ vertical line in the phase diagram of Figure~\ref{vesicle_phase_diagram}c. When $\kappa_1$ grows above $\kappa_1^*$ as given in Equation~\eqref{kappa1star}, the~liquid phase becomes a structured disordered one where density fluctuations correspond to thicker transient domains. Increasing further $\kappa_1$ leads to even more structured mesophases, as in the strong segregation~case.

As far as spontaneous curvature is concerned, a temperature $T^*(C_1)$ has been defined in Section~\ref{curv:comp:cp}. Below $T^*$  and for large enough $C_1$, mesophases are favored by spontaneous curvature~as~in the strong segregation limit. Above $T^*$, the structured disordered phase now corresponds to small transient domains of the spontaneously curved phase. If only spontaneous curvature is taken into account and no dependence of $\kappa$ with $\phi$ is considered, one must focus on the $\kappa_1=0$ horizontal line in Figure~\ref{vesicle_phase_diagram}c. When $C_1$ grows above a certain threshold on the order of $C_1^*$, the~liquid phase becomes a structured disordered one. In the planar geometry, Figure~\ref{curvature_phase_diagram}b also reveals that the macrophase below $T^*$ is destabilized by spontaneous curvature, leading again to a modulated mesophase. The same observation can be made in Figure~\ref{curvature_phase_diagram}a in the asymmetric case (large values of $m_0 \gtrsim 8$ for the parameter values of the figure).

\bigskip

\section{Active and Out-of-Equilibrium Processes}
\label{active:memb}

Alternatively to the approaches in thermodynamic equilibrium  discussed so far, some works have discussed for a dozen years~\cite{Jacobson2007,Hancock2006} the possibility that out-from-equilibrium, active processes might explain the observed domain finite size. Generally speaking, the mechanisms at play appeal either to the active remodeling of the cortical cytoskeleton (which is coupled to the membrane~\cite{Lenne2009}) or to the rapid traffic of membrane components. Traffic occurs to and from the membrane, either by exchange of monomers or by endocytosis and exocytosis of membrane patches through vesicles of $\sim 100$~nm in diameter~\cite{kobayashi1998}, with a surface of $10^4$ to $10^5$~nm$^2$. This active exchange of membrane material is often called ``membrane recycling''. The relationship between domain size and endocytosis rates has been explored experimentally in at least one biological system~\cite{Quang2013}: the data suggest that blocking the endocytic pathway leads to an increased number of large clusters of E-cadherins (containing more than $\sim100$ copies), supporting the idea that the growth of large domains is impeded by recycling. It was concluded by the authors of reference~~\cite{Quang2013} that ``endocytosis targets large E-cadherins clusters, while not perturbing the mechanisms of clustering at scales smaller than 100'' copies.

\begin{figure}[h]
\begin{center}
\begin{tabular}{cc}
\parbox{7cm}{
\includegraphics*[width=7cm]{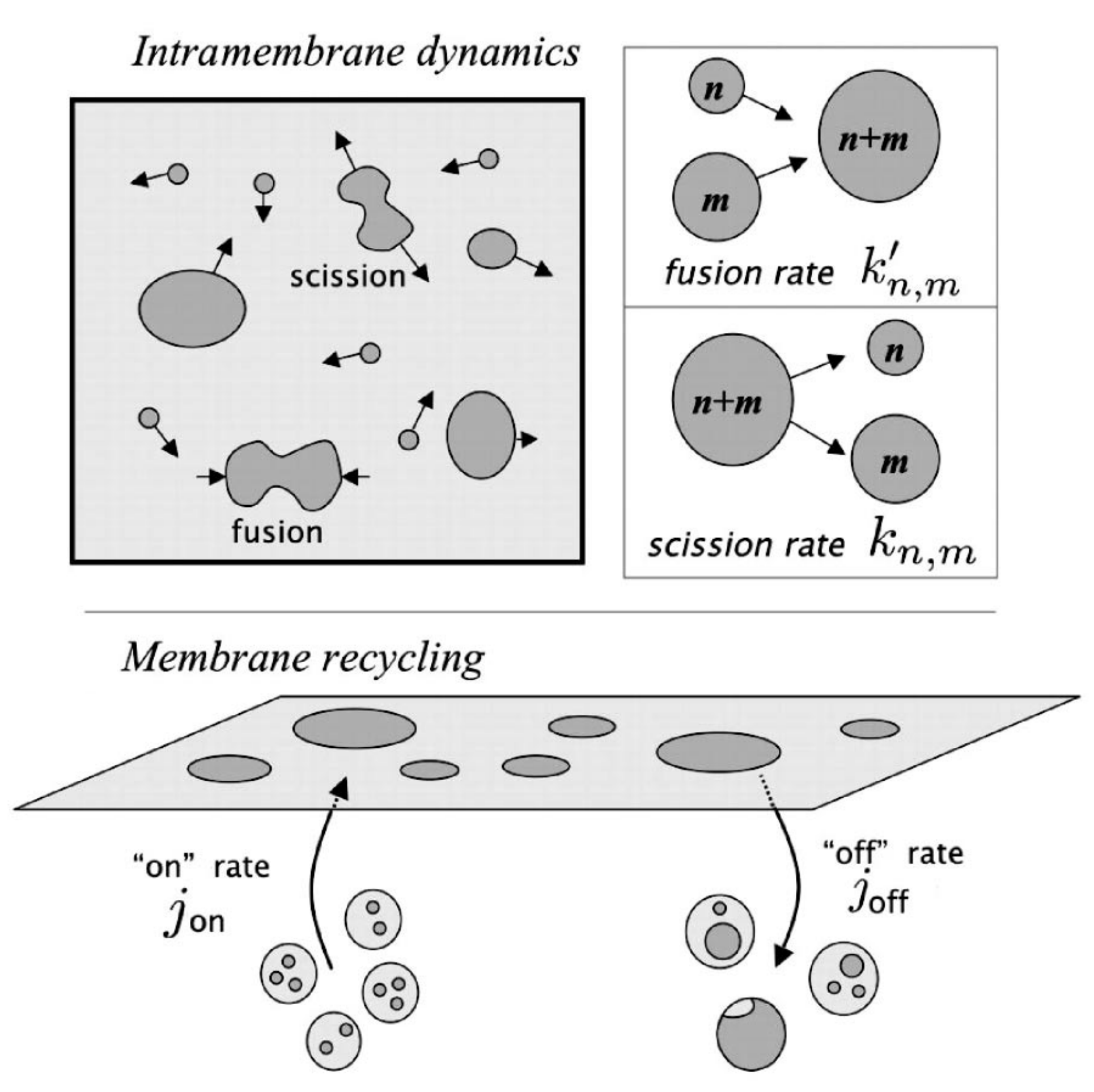}}
& \quad \parbox{9cm}{\includegraphics*[width=7cm]{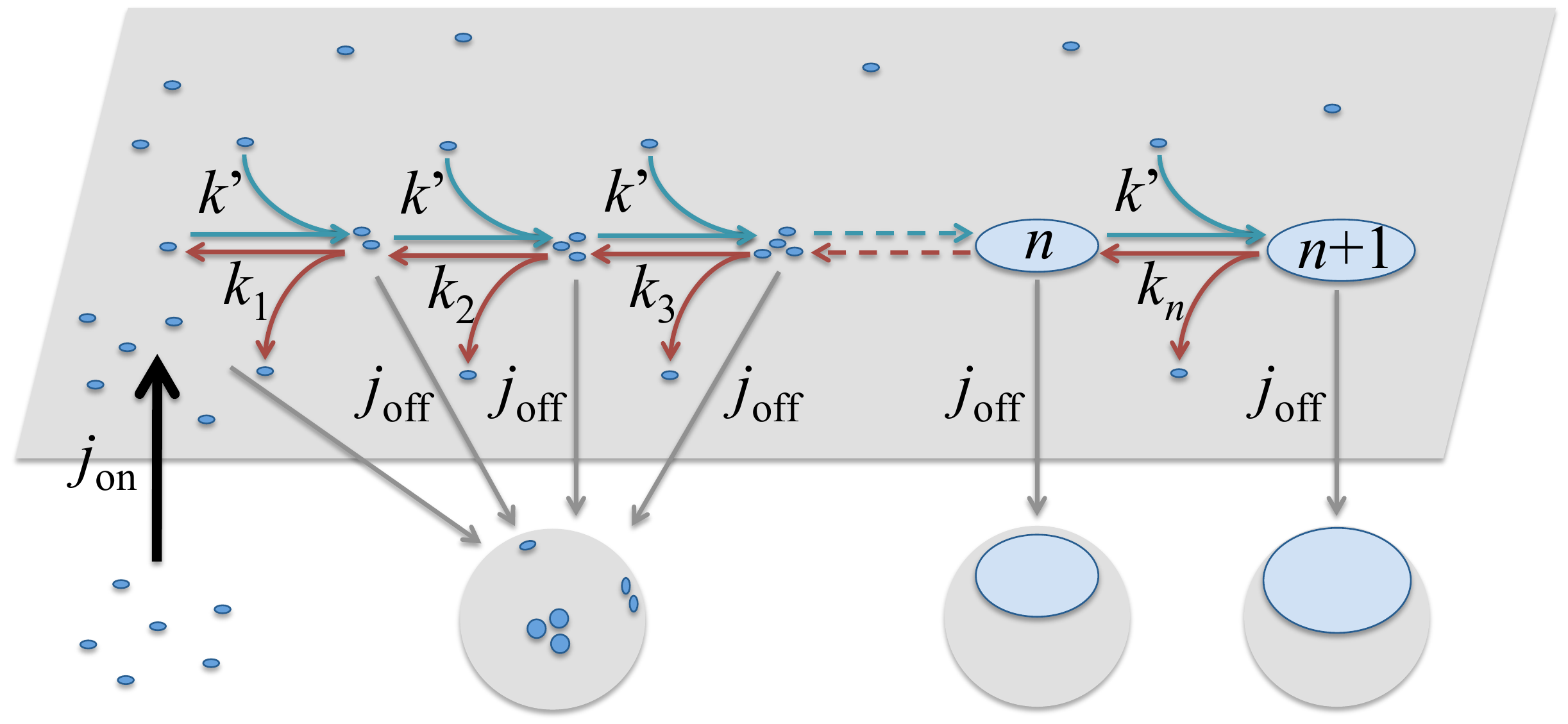}}
\end{tabular}
\end{center}
\ \hfill \hspace{1cm}  (\textbf{a}) \hspace{8cm} (\textbf{b}) \hspace{2cm} \hfill \
\caption{Examples of recycling schemes. (\textbf{a})~In the membrane, domains can undergo scission or fusion, whatever their size. Reproduced from~\cite{Turner2005}, with permission from the American Physical Society, Copyright 2005. (\textbf{b})~In this alternative scheme, clusters can only gain or lose proteins by exchange of monomers with the surrounding membrane (Ostwald ripening) because scission and fusion events are assumed to be rare. Reproduced from~\cite{Berger2016}, with permission from the American Chemical Society, Copyright 2016.  In both examples, monomers are injected into the membrane from the cytosol at a rate $j_{\rm on}$, either by exocytosis in (\textbf{a}) or by a monomer flux from the cytosol in (\textbf{b}). Multimers are internalized through endocytosis with a rate $j_{\rm off}$ that is independent of their size. }
\label{Berger:Fig}
\end{figure}

Experiments also indicate that the typical amount of plasma membrane endocytosed per hour is equivalent to the total surface of the cell plasma membrane of a fibroblast~\cite{Steinman1983}. Faster recycling times for membrane components have even been measured, on the order of 10~min or even shorter (see~\cite{Hao2000} and references therein).  As illustrated in Figure~\ref{Berger:Fig}, recycling enters in competition with the growth of domains, as driven either by domain diffusion followed by their coalescence, or by Ostwald ripening~\cite{Bray1994,Rosetti2017} (see below). Here the domain coalescence is still driven by a finite line tension below $T_d$ which tends to minimize the interface length between Lo and Ld regions. Note that as in Abney and Scalettar's work~\cite{Abney1995}, Gheber and Edidin had already anticipated such a mechanism in 1999, above $T_c$ however~\cite{Gheber1999}. In their model, an additional meshgrid of obstacles hinders diffusion, mimicking the presence of the cortical actin cytoskeleton, and ensures a sufficient lifetime of domains by preventing the too fast diffusion of membrane molecules. Note also that if the transport between the membrane and the cytosol is passive, without any energy consumption, the membrane is actually in equilibrium, in contact with a molecule reservoir. One recovers the equilibrium case discussed in the previous sections, but now described in the grand-canonical ensemble~\cite{Foret2012}.

Active membrane recycling can fracture large assemblies or domains, the growth of which (below the demixing temperature) is thus arrested at a finite typical radius $L^*$ in the steady state~\cite{Foret2005,Turner2005,Fan2008,Gomez2009}. As in the equilibrium case, these models have first been designed to account for the finite size of Lo lipid domains, however, at this stage they are sufficiently general to be applicable to protein clusters as well. Both cases will be treated in parallel in this section.

\subsection{Models}
\label{models}

A basic but illuminating scaling argument was proposed by Foret in 2005~\cite{Foret2005}.  When quenching a homogeneous lipid mixture below the demixing temperature, it phase-separates. Under certain realistic conditions, the typical domain size coarsens as $L(t) \propto (D\, a\, t)^{1/3}$~\cite{Bray1994}, where $D$ is the two-dimensional diffusion coefficient, and $a$ is the typical lipid (or protein) lateral size, in the nanometer range. Additionally, traffic to and from the membrane sets a typical recycling timescale $\tau$, as discussed~above. Recycling prevents equilibration beyond the timescale $\tau$, and the coarsening stops at a typical domain radius $L^* \propto (D\, a\, \tau)^{1/3}$. Assuming $D \lesssim 0.1~\mu$m$^2$/s for cell membrane constituents~\cite{Phillips2015}, $a$ smaller than few nm for a typical membrane lipid or protein, $\tau \leq 1$~hour and ignoring numerical prefactors in the expression of $L^*$, one gets $L^*  \lesssim 1~\mu$m. This rough estimate is encouraging, just one order of magnitude larger than typical nanodomain sizes.  

This argument has been completed and confirmed by several theoretical studies in the last decade. They dwell on a variety of theoretical techniques, such as linear stability analysis in the Fourier space~\cite{Seul1995,Foret2005,Gomez2009,Gomez2008}, or numerical and/or analytical resolutions of stochastic equations belonging to two principal categories:
\begin{itemize}
\item[(i)] Master Smoluchowski's coagulation equation~\cite{Turner2005,Foret2012,Berger2016,Vagne2015}. For example, Turner et al.~\cite{Turner2005}  studied the coagulation equation 
\begin{equation}
\frac{{\rm d} c_n}{{\rm d} t} = \zeta(n) + \sum_{m=1}^{\infty}  k_{n,m} c_{n+m} - k'_{n,m} c_{n}c_m + \frac12  \sum_{m=1}^{n-1}  k'_{m,n-m} c_{n-m}c_m - k_{m,n-m} c_{n},
\end{equation}
where $c_n$ is the area fraction of domains or clusters containing $n$ monomers and the constants $k_{n,m}$ and $k_{n,m}'$ are scission and fusion rates, respectively. Finally, $\zeta(n)$ controls recycling. Both cases~in Figure~\ref{Berger:Fig}a,b correspond to $\zeta(n)=j_{\rm on} \delta_{n,1} - j_{\rm off} c_n$ because only monomers are injected in the membrane with a rate $j_{\rm on}$ while domains are recycled to the cytosol with a rate  $j_{\rm off}$ independent of their size $n$.  
\item[(ii)] Non-linear reaction-diffusion equations~\cite{Foret2005,Fan2008,Gomez2009,Gomez2008,fan2010c}, which can also be seen as Cahn-Hilliard equations~\cite{Chaikin,Bray1994} suitably modified to take recycling into effect. For instance, in~Ref~\cite{Foret2005}, the~Cahn-Hilliard equation
\begin{equation}
\label{eq24}
\frac{\partial \phi}{\partial t} = - \frac{\phi-\bar \phi}{\tau} + A \nabla^2 \frac{\partial w}{\partial \phi} 
\end{equation}
is used. The order parameter $\phi(\mathbf{r},t)$ again measures the local concentration of one phase (say the Lo phase), $\bar \phi$ is its equilibrium value, $\tau$ is the recycling rate as discussed above, \mbox{and $A$ is a} kinetic constant proportional to the lipid diffusion coefficient. The function $w(\phi)$ is the Landau-Ginzburg expansion of a membrane patch free-energy, at the relevant order in powers of the order parameter~$\phi$, as already evoked in Section~\ref{equil}. Recycling is embedded in the first term of the~r.h.s, while the second term describes the diffusive transport of the conserved order parameter~$\phi$. 
\end{itemize}
 
\medskip

The different models can be further categorized as follows (see also references~\cite{Foret2012,Fan2010a} for alternative~discussions):
\begin{enumerate}
\item The off-rates (from the membrane to the cytosol) can be size-dependent~\cite{Quang2013} or not~\cite{Turner2005,Berger2016}. In the former case, it means for example that endocytosis is able to extract patches from the membrane with a limited size set by the endocytosed vesicle typical size~\cite{Vagne2015}. A ``recycling correlation length'' can also be introduced in the modified Cahn-Hilliard equations, mimicking the spatial range of recycling processes, i.e., the typical size of membrane patches recycled through vesicle traffic~\cite{Fan2008,fan2010c}. In the models of Refs.~\cite{Foret2005,Turner2005,Foret2012,Gomez2009,Gomez2008}, only monomers are locally extracted from the membrane.
\item The on-rates (from the cytosol to the membrane) are also size-dependent. Several models only inject monomers or tiny domains in the membrane~\cite{Foret2005,Turner2005,Quang2013,Foret2012,Berger2016,Gomez2009,Gomez2008,Vagne2015} because they do not assume any pre-order in the exocytosed patches or because they assume direct exchange of monomers from the cytosol to the membrane, e.g., for peripheral proteins. Indeed, Foret argues that the traffic should be modeled differently for peripheral and transmembrane proteins~\cite{Foret2012}, because the former are preferentially exchanged as monomers between the cytosol and the membrane, while the latter preferentially escape and join the membrane by endo- and exocytosis, respectively. Another approach assumes that domains with a characteristic size are directly injected in the membrane~\cite{Fan2008,fan2010c}. 
\item Inside the membrane, two mechanisms control the domain dynamics: either the domains principally exchange matter through Ostwald ripening (exchange of monomers via the surrounding dilute ``gas'' phase~\cite{Bray1994,Rosetti2017}, as illustrated in Figure~\ref{Berger:Fig}b), see references~\cite{Foret2005,Foret2012,Berger2016,Fan2008,Gomez2009,Gomez2008,fan2010c}, or through domain scission or fusion events, for all sizes~\cite{Turner2005,Quang2013} (Figure~\ref{Berger:Fig}a).
\end{enumerate}

To finish with, an alternative viewpoint has considered intermittently active proteins, which~undergo conformational transitions driven by external energy sources of any form (such~as light for photoreceptors, or ATP for other active proteins)~\cite{Prost1996,Sabra1998}. The model assumes that the protein affinity for the Lo or Ld phases depends on its conformation, for example because of hydrophobic mismatch energy. Otherwise macroscopic domains are then split into finite-size ones. \mbox{Indeed, large~protein} assemblies are destabilized because protein changing conformation continually leave them, as it is the case with recycling above. Accordingly, the typical domain size is demonstrated to decrease with enhanced protein activity.

\subsection{Results and Prospects}
\label{prospects}

\begin{figure}[h]
\begin{center}
\includegraphics*[height=6.5cm]{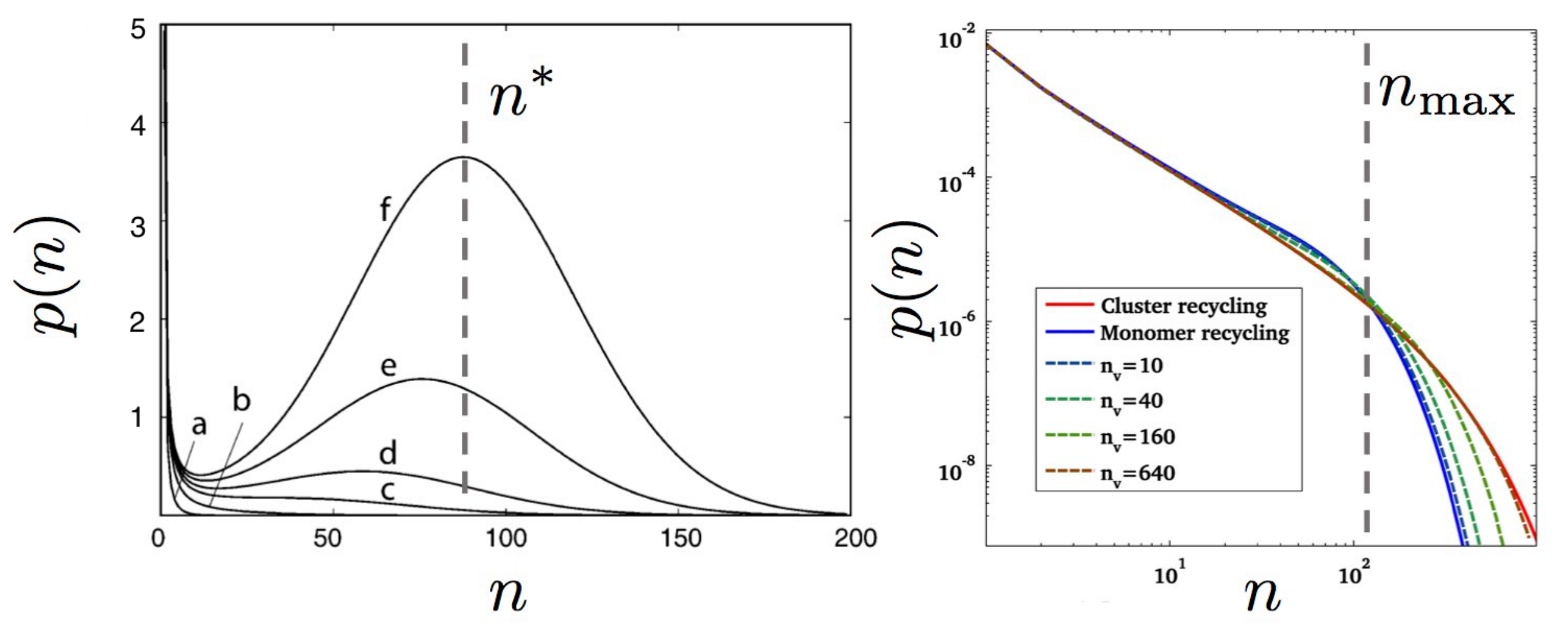}  
\end{center}
\ \hfill \hspace{2cm}  (\textbf{a}) \hspace{7cm} (\textbf{b})  \hfill \
\caption{Examples of steady-state domain-size distributions $p(n)$. Here the domains are protein clusters. In these figures, their size is measured as the number $n$ of proteins they contain, proportional to $L^2$ if $L$ is the cluster radius. The units on the vertical axis are arbitrary. (\textbf{a})~Bimodal distributions, where small oligomers coexist with domains of typical size $n^{*}$ (vertical dashed line for curve f). When going from curve ``a'' to curve ``f'', the number of monomers in the membrane increases because more and more monomers are injected in the out-of-equilibrium membrane from the cytosol. Above a critical value, multimers nucleate and the distribution becomes bimodal as in curves ``d'', ``e'' and ``f''. The limit between monomodal and bimodal distributions is curve ``c''. Adapted from~\cite{Foret2012}. (\textbf{b})~Power-law distributions $p(n) \propto n^{-3/2}$ with a cut-off $n_{\rm max}$  (vertical dashed line) above which $p(k)$ decreases exponentially. Here the coalescence rate between two clusters is independent of their size. Clusters smaller than a size $n_{\rm v}$ are recycled entirely as a whole, while larger clusters are fragmented and lose an area $n_{\rm v}$ during a recycling event. The ``whole cluster recycling'' limit corresponds to $n_{\rm v} \rightarrow \infty$ and the ``monomer recycling'' limit to $n_{\rm v} = 1$. Adapted from~\cite{Vagne2015}. For both figures, see the cited references for more details on the recycling dynamics. 
\label{distribs}}
\end{figure}

These models agree on the existence of finite-size domains in the steady state instead of the macrophase that would prevail in equilibrium. However, they can lead to quantitatively different predictions concerning the steady-state distributions $p(L)$ of domain sizes, depending on the recycling scheme details. As illustrated in Figure~\ref{distribs}, the different approaches can predict 
\begin{enumerate}
\item Either bimodal size distributions~\cite{Foret2005,Foret2012,Berger2016,Gomez2009,Gomez2008}, where monomers or small oligomers (the~``gas'' phase already evoked in Section~\ref{cl:ph}) coexist with larger domains (the ``condensed'' phase) with a typical size $L^{*}$;
\item Or large, power-law distributions $p(L) \propto L^{-\delta}$, with a large domain-size cut-off above a limiting upper size $L_{\rm max}$, above which domains become extremely rare~\cite{Turner2005,Quang2013,Vagne2015}.
\end{enumerate}

All models predicting a bimodal distribution inject only monomers in the membrane from the cytosol and are based on the Ostwald ripening dynamics inside the membrane (exchange of monomers between clusters). These two prerequisites will have to be confirmed with a solid justification~in future~studies. In contrast, the way off-rates depend on the domain size does not seem determinant in setting the shape of $p(L)$. 

Note also that Smoluchowski's coagulation-equations where clusters of any size can coagulate~or split into smaller ones have been known for long to lead to power-law distributions with a cut-off~in quite generic circumstances~\cite{Family1986}. The fact that clusters of any size can coagulate or not depend~on whether they feel mutual repulsion or not. Mutual repulsion can be attributed to membrane spontaneous curvature inside clusters~\cite{Berger2016}. The fact that clusters can split into smaller multimers depends on the line tension $\lambda$. If  $\lambda$ is too large, creating two smaller new clusters is quite expensive~in terms of interfacial energy and Ostwald ripening
is favored. Hence the shape of the distribution $p(L)$ certainly strongly depends on the way the different membrane constituents interact at short and longer~range.

To conclude this section on active out-of-equilibrium processes, it is worth emphasizing that experiments are absolutely required to clarify the rich but somewhat involved landscape of the theoretical predictions that we have described. To our knowledge, the experimental approach proposed in Ref.~\cite{Quang2013} is the unique one to bring insight onto the effect of membrane recycling rates on membrane domain sizes. Experimental are thus wanted to assert (or not) the relevance of the sometimes contradictory assumptions on which the different models discussed above rest. 

\section{General Discussion and Conclusions} 

\subsection{A Variety of Mechanisms in Equilibrium in the Strong Segregation Limit}

We have focussed in this review on biphasic fluctuating membranes that phase-separate in two phases at sufficiently low temperature. We have named these phases Lo and Ld even though the formalism at play can in principle be extended to a wider context, in particular to protein clusters in an homogenous lipid mattress. We have classified the various mechanisms involved in the stabilization~of mesophases and we have argued that {many of them}  dwell on the broken symmetry between the inner and outer (or upper and lower) leaflets. 

Even in the case of simplified lipid bilayer mixtures, where only three or four lipid species coexist, the phenomenology is already dramatically rich, as exemplified by Feigenson and collaborators' studies. Here we discuss the different mechanisms presented from the perspective of their relevance in the cell biology context. Which mechanism(s) is/are likely to be at play in a real cell to promote the observed nanodomains having a 10 to 200~nm diameter, commonly called ``rafts'' in the cell biology community? 

Buckling induced by a strong line tension (see Section~\ref{buckling}), as first deciphered by Lipowsky and collaborators in the 1990's~\cite{Lipowsky1992,julicher}, has been shown to lead to \emph{micrometric} 
domains in the metastable state where buckled domains feel mutual repulsion.  We have seen that the line tension required~to promote buckling of Lo domains is $\lambda>8 \kappa_{\rm Lo}/L$ for domains of radius $L$ in taut membranes. For~$ \kappa_{\rm Lo} \geq 100$~$k_{\rm B}T$ as measured experimentally, buckling of domains of radius $<100$~nm require $\lambda> 30$~pN, an order of magnitude larger than measured line tensions. The situation is even worse for 10-nm nanodomains, the admitted lower bound of raft sizes. These values seem to exclude buckling~as~a generic realistic mechanism in the cell biology context. In addition, measured critical temperatures $T_c$ lie below the physiological temperature, in the $T_c=20$ to 30 $^\circ$C range, even in giant plasma membrane vesicles~\cite{veatch2}. It ensues that Lo and Ld phases are probably not  separated in living~cells. \mbox{However, this last} argument must be taken with care because many membrane constituents are missing in model membranes. They might shift the actual critical temperature to higher values.

In the phenomenon studied by Lipowski's and Feigenson's groups (see Section~\ref{carapace}), the more rigid Lo phase ``carapace'' must be split into smaller domains to accommodate it to the global spherical shape of the GUV~\cite{Gutlederer2009,Konyakhina2011}. With the typical bending moduli of Lo phases given above, and GUVs with diameter $R> 10$~$\mu$m, the domains are observable by classical optical microscopy and are again~on the micrometer scale. Getting much smaller nanodomains requires to increase the background mean curvature $1/R$. This can be achieved on highly curved cell substructures \mbox{such as microvili}, \mbox{small~cellular vesicles}, dendrites or axons, but it is unlikely on flatter membrane regions where nanodomains are also observed. The same remark as above concerning the critical temperature holds.

In contrast, domains ensuing from the spontaneous curvature $C_{\rm sp} \neq 0$ induced by some of their constituents can readily lead to nanodomains for realistic parameter values because they make too large domains unstable, as discussed in Section~\ref{cl:ph}. We have explained how a large variety of mechanisms can promote membrane spontaneous curvature by playing on the asymmetry between both leaflets. Without excluding that the two previous mechanisms are also more marginally at play, spontaneous curvature is probably the most efficient mechanism in the strong segregation limit. 

Linactants are also known to reduce the line tension $\lambda$ and to promote metastable micro-emulsions~by localizing at the interface between coexisting phases, as surfactants do in many soft condensed matter phenomena~\cite{SafranBook,Honerkamp2009}. This role has been hypothesized for cholesterol or asymmetric saturated/monounsaturated lipids, like POPC and 1-stearoyl-2-oleoyl-sn-glycero-3-phosphocholine (SOPC) 
, although this last possibility has been contested (see references~\cite{Heberle2010,Heberle2013,Usery2017,brewster} and references therein). From a practical point of view, reducing the line tension $\lambda$ with the help of linactants is comparable to driving the system closer to a critical point where $\lambda$ vanishes~\cite{Schmid2017}.

\subsection{Nanodomains and Critical Density Fluctuations}
\label{disc:fluct}

The previous section dealt with the strong-segregation limit. In Section~\ref{cl:strong}, we have underlined that the density fluctuations in the weak-segregation limit are quite specific when membrane shape and composition are coupled. In absence of coupling, biphasic systems close to criticality belong to the Ising universality class and the characteristic fluctuation length-scale is set by the Ornstein-Zernike correlation length $\xi_{\rm OZ}$ which diverges at the critical point and remains large in its vicinity~\cite{Chaikin,Mouritsen2005,Honerkamp2009}. Above $T_c$, the domain-size distribution decays with a power-law up to a maximum size set by $\xi_{\rm OZ}$~\cite{Cornell2018}. By contrast, when coupling is switched on, the nanodomain size in the Structured Disordered phases~is set by the Helfrich correlation length $\xi$ which has no reason to be comparable to $\xi_{\rm OZ}$, and the domain-size distribution is peaked at $\xi$ where the structure factor has a maximum. 

Since critical density fluctuations are good candidates for the nanometric rafts predicted by biochemical techniques~\cite{Komura2014,Destain2016,Schmid2017}, they have been probed by a large variety of experimental approaches. Indeed $\sim10$~nm nanodomains in ternary lipid mixtures can be observed by FRET and electron spin resonance~\cite{Heberle2010,Feigenson2001}, SANS~\cite{Heberle2013} or AFM 
~\cite{Connell2013,Ho2016}. However, the domain size distribution cannot be determined through these experimental techniques and only conjectures can be made concerning their precise nature. It cannot easily be ascertained whether they are just the manifestation of density fluctuations in the vicinity of a critical point or correspond to the structured disordered phases evoked~above. 

To clarify this issue, Veatch and collaborators observed by fluorescence microscopy giant plasma membrane vesicles (GPMVs) directly isolated from cells~\cite{veatch2}. As compared to the model three- or four-component membranes discussed so far, GPMVs retain many of the original plasma membrane constituents, lipids and proteins, even though some up-down asymmetry is likely lost during the GPMV formation~\cite{Honerkamp2009}. They also separate into Lo and Ld phases at low temperature. Even though the resolution is not as good as in AFM experiments, fluctuations above $T_c$ could be analyzed at scales $\geq 150$~nm, i.e., above the Rayleigh limit. Their observations were qualitatively consistent with all GPMVs having near-critical compositions, suggesting that plasma membranes have compositions that are naturally designed to reside near a miscibility critical point in vivo. Thermodynamic quantities such as the correlation length above $T_c$, the line tension below $T_c$, the compressibility and the diffusion coefficient could be inferred close to a critical point. In spite of important error bars, the authors concluded that all of them indicated that the system belongs to the 2D Ising universality class, \mbox{i.e., that of an} ordinary binary mixture close to its critical point. A similar study was conducted~on GUVs made of a ternary lipid mixture close to its critical point and the authors were led to similar conclusions~\cite{Honerkamp2008}. However, in such studies subject to Rayleigh limit, the presence of a peak in the structure factor at short wave-lengths $ \ll 150$~nm, as predicted by coupled theories, cannot be excluded. 

Connell and collaborators circumvented this limitation in 2013 by examining very accurately the bilayer composition fluctuations with an AFM~\cite{Connell2013}. Lo and Ld phases can easily be distinguished because they do not have the same thicknesses. Importantly, the authors observed that ``The region~of critically fluctuating 10--100 nm nanodomains has been found to extend a considerable distance above $T_c$ to temperatures within the biological range, and seem to be an ideal candidate for the actual structure of lipid rafts in cell membranes.'' Domains of size $\geq10$~nm could be observed at temperatures up to 7 $^\circ$C higher than $T_c$. They also concluded that supported DOPC/SM/cholesterol bilayers belong to the 2D Ising universality class because the measured structure factors have the expected shape and the critical exponent associated with the correlation length is close to the 2D Ising-class value, $\nu=1$. However, the membrane observed by AFM are supported, which suppresses wave-length fluctuations above some cut-off depending on the strength of the substrate-membrane attraction, and potentially inhibits the effects of shape-composition coupling. These experiments have not definitely ruled out the existence of structured disordered phases either. 

To finish with, a potentially important structural difference between domains below the demixing temperature $T_d$ and density fluctuations close to the critical point has recently been suggested through coarse-grained molecular dynamics~\cite{Baoukina2017}. Well below $T_d$, there is a strong overlap between Lo domains~in opposed leaflets (at least in planar geometry), which we have called registered domains~above. In contrast, the overlap is negligible in the case of density fluctuations (anti-registration). This difference could be fully relevant from a biological perspective.

\subsection{Needed Theoretical Clarifications}
\label{needed:theory}

When observing 2D patterned morphologies, it appears that in some circumstances, domains are more or less roundish, like bubbles, whereas they become more elongated, leading to interdigitated, labyrinthine structures, in other situations (see, e.g., Figure~\ref{Feigenson:Fig}c). They can even adopt a relatively regular striped geometry in some cases. This issue was already examined in the articles of Andelman and collaborators~\cite{Seul1995,leibler2} because shape instabilities have been studied for long, \mbox{back~to the 1970's}, for~example in ferrofluid films in a magnetic field or polar phospholipid Langmuir films~at suitably chosen temperatures and pH. Fluorescence microscopy imaging on vesicles made of a SM/DOPC/cholesterol mixture below the phase-separation temperature have shown that there is a transition from stripes to bubbles when the reduced volume $v$ grows and approaches 1, presumably because of increasing surface tension~\cite{Rozovsky2005}, as later confirmed by field-theoretic arguments based~on the Hamiltonian~\eqref{couplingLA} and the single-mode and weak-segregation limit approximations in planar geometry~\cite{Komura2006}, {and much more recently by new experiments~\cite{Cornell2018}}. By using the same kind of arguments, a transition from bubbles to stripes is also predicted when increasing the concentration of the minority phase at fixed thermodynamic conditions or lowering the temperature~\cite{leibler2,Konyakhina2011,Harden1994,Harden2005}. However, as pointed out in Ref.~\cite{Komura2006}, a consensual understanding of the relative stabilities of the different morphologies remains necessary. Experiments with controlled surface tension are also still lacking to guide theoretical developments. In the even more complex cell biology contexts, experimental data tend to confirm that increasing the protein concentration leads to more elongated protein clusters~\cite{Merklinger2017} (see Figures~\ref{Feigenson:Fig} and 5S of this reference). More systematic experimental studies in connection with theoretical works are also needed. 

The understanding of the role of electrostatics in shaping the membrane, even though already considered in the 1990's~\cite{Guttman1993}, also remains lacunar. As already indicated in Section~\ref{carapace}, reference~\cite{Amazon2014}, followed by reference~\cite{Usery2017}, studied numerically how dipole-dipole or charge-charge repulsion between~lipids, having a nanometric range because of screening, competes with the shorter-range attraction between lipids of the same phase. As expected~\cite{Seul1995}, competing interactions are found~in this work to lead to small domains, of size $\sim 10$~nm, provided that line-tension is not too large (in which case macrophase separation occurs instead). 
As discussed in the review by Schmid~\cite{Schmid2017}, the uncertainty of how to finely model membrane leads to contradictory predictions concerning the resulting domain~size. In addition, we also remarked above that models embracing both nanoscopic and micrometric length-scales remain to be developed. {Very few works have tackled} them~on~an equal footing~so~far, probably because the level of coarse-graining cannot be the same for the different scales {which makes numerical studies difficult}. A multi-scale approach is needed. If both {electrostatic} lipid-lipid repulsion~at nanoscopic scale and a mechanism as described above leading~to larger domains~are~present, does one expect a hierarchy of domains, with nanoscopic domains embedded~in microscopic ones? {Such a scenario has been proposed on analytical grounds in~\cite{liu2005} where, again~appealing to the Gaussian approximation of a 3-field theory,  a region of the parameter space was found where nanometric ``raft'' domains dwell in larger, micrometric ``nonraft''~ones, themselves distinct from the ``background'' components.} Such hierarchy has also been suggested by different experimental techniques~\cite{Murase2004,Saka2014,Mocsar2016}, but its {connection with theoretical findings will have to be confirmed in future studied.}

Furthermore, we have evoked that some studies presented in this Review include Gaussian curvature and the saddle-splay modulus $\kappa_G$ in membrane modeling~\cite{Semrau2009,Gutlederer2009,Hu2011}. It is generally assumed that  $\kappa_G \approx - \kappa$ to ensure membrane stability~\cite{Gutlederer2009}. However, in most analytical and numerical works, the corresponding second term in Equation~\eqref{Helf} is neglected, by arguing that at fixed topology, its~integral over the whole membrane surface is an invariant due to the Gauss-Bonnet theorem. But if $\kappa_G$ turns out~to depend significantly on the lipid phase, this argument fails and a more rigorous \mbox{approach is needed}. J\"ulicher and Lipowsky have tackled this question in 1996~\cite{julicher}, using the fact that when $\kappa_{G, {\rm Lo}} \neq  \kappa_{G, {\rm Ld}}$, the integral over the whole surface can be simplified, still owing to the Gauss-Bonnet theorem. \mbox{It is proportional} to $\kappa_{G, {\rm Lo}} - \kappa_{G, {\rm Ld}}$ and can be written as the integral of the geodesic curvature along the interface between both phases, see Equation~(\ref{eq24}) of reference~\cite{julicher}. It can potentially favor a longer interface and thus a mesophase~\cite{Adkins2017}. However, J\"ulicher and Lipowsky's study was only dedicated to the shape of vesicles undergoing macrophase separation, and not to the stability of mesophases. To our knowledge, a systematic study of the interplay between the different contributions to the membrane energy remains to be done to ascertain the precise role played by the Gaussian curvature and clarify when it can be neglected or not.

\subsection{Needed New Experiments}
\label{needed:exp}

In Section~\ref{active:memb}, we have reviewed the different mechanisms proposed in the literature to take active membrane recycling into account. At this stage, there is no definitive reason to favor mechanisms in equilibrium with respect to active ones. However, nanodomains can be observed in membrane fragments peeled off from cells long before their observation. This tends to prove that active processes are not absolutely required to maintain nanodomains. For example, retina cell fragments from dead human and murine corpses have been observed by AFM. They presented very well characterized rhodopsin nanodomains~\cite{Whited2015}. This does not preclude that active processes can come in addition to equilibrium arguments and reorganize the mesophases. Berger and her coworkers have recently combined such out-from-equilibrium arguments and a competing-interaction model as discussed in Section~\ref{cl:ph} to explore how recycling modifies the equilibrium domain sizes~\cite{Berger2016} (see Figure~\ref{Berger:Fig}b above). The main finding of this analytical work is that when taking recycling into account, the typical cluster size at steady state increases very slowly, logarithmically with the recycling rate. Using physically realistic model parameters, the predicted two- or three- fold increase of the typical cluster size due to recycling in living cells is likely experimentally measurable with the help of super-resolution microscopy, even though such experiments might be delicate to implement because the typical cluster size depends on many other parameters. In addition, in this model and contrary to others, an {\em increase} of the typical clusters size is predicted with increased recycling at fixed protein density, because only monomers are injected in the membrane and only monomers are exchanged between clusters (what we have called Ostwald ripening~\cite{Rosetti2017}). However, as already evoked in Section~\ref{prospects}, cluster- or domain-size distribution shapes (Figure~\ref{distribs}) should help to discriminate between the different propositions based on theoretical considerations. They will have to be confronted to high-precision experiments in the future, thus enabling biologists and biophysicists to identify the mechanisms that are actually at play in living~cells.

{Some experimental studies~\cite{Feigenson2001,Goh2013} have suggested a transition between two distinct modulated phases, one with a nanoscopic domain length-scale (possibly critical density fluctuations), and one with much larger domains. Does the apparently discontinuous character of the phase diagram just come from the diversity of the experimental techniques used to probe the different regimes of parameters? This point will have to be clarified in future studies. It is closely akin the problem raised in the previous section about the domains predicted at different scales by different theoretical approaches.}

In this work, we have considered Lo domains and protein domains on an equal footing, because they share many common characteristics from a theoretical perspective. Real rafts in cells are assumed to gather together specific proteins in a Lo phase nanodomain. In the specialized literature, deciphering the respective roles of lipids and proteins in the stabilization of rafts is sometimes controversial. Does lipidic mesophase-separation precede targeting of protein to rafts or do proteins gather and bring their specific lipid annuli with them, thus enriching nanodomains in some lipid species? This somewhat sterile debate seems useless to us because a nanodomain is stabilized by the aggregation of all its constituents, lipids and proteins, that feel mutual attraction at short range, as discussed in detail in this Review. However, the important question is to understand what sets their finite, nanoscopic size. Imbedding membrane proteins to three- or four-component lipid GUVs will probably constitute a step further towards the understanding of how lipids and proteins collaborate in real cells.

\subsection{Concluding Remarks}

Some major pieces of the jigsaw puzzle have been identified, but the subtle way in which they assemble is still the object of passionate debates. We have argued that the best candidates in equilibrium to be the so-called rafts in real cells are: (i)~nanodomains induced by the curvature of some of their constituents, in the strong-segregation limit,  as discussed in depth in Section~\ref{cl:ph}; or (ii) composition fluctuations in the vicinity of a miscibility critical point, as described in Sections~\ref{aboveTm}~and \ref{disc:fluct}. Both mechanisms likely coexist in real cells. However, a somewhat confusing situation prevails in the biophysics literature, in particular because the physical concepts are not always used in their strict domain of validity. As pointed out by Veatch and Keller, ``it can be confusing to use the language of phase separation when describing small features [composition fluctuations] because they differ in many ways from large domains [meso- or macro-phases]''~\cite{veatch1}. In addition, for both kinds of domains, their size distributions are potentially modified by active, out-of-equilibrium phenomena (see Section~\ref{needed:exp}).

A way to disentangle further this long-standing issue might be to characterize better domain lifetimes. Nanodomains must have sufficiently long lifetimes to allow their constituents to interact~\cite{Heberle2010}. Measured lifetimes of composition fluctuations are on the order of 1~s, due to their rapid destruction by diffusion~\cite{veatch2}. Is it sufficient to enable nanodomains to play their biological role of micro-reactor? To our mind, this is one of the major issues that ought to be challenged in the forthcoming years.  



\subsection*{acknowledgments}
We are indebted to Guillaume Gueguen, Evert Haanappel and Laurence Salom\'e for fruitful discussions and sound advice.
 


\end{document}